\documentclass[12pt]{article}

\pdfoutput=1

\usepackage{graphicx}
\usepackage{hyperref}
\usepackage{amsmath}
\usepackage[table]{xcolor}
\usepackage{color}
\begin{document}

\hoffset = -0.3truecm
\voffset = -1.1truecm

\title{\bf MAP, MAC, and Vortex-rings Configurations in the Weinberg-Salam Model\footnote{To be submitted for publication}}

\author{
{\bf Rosy Teh\footnote{Permanent address from 30th March 2015 onwards: 48, Halaman Bukit Gambir 5, 11700 Gelugor, Penang, Malaysia; email: rosyteh@gmail.com}}, 
{\bf Ban-Loong Ng and Khai-Ming Wong}\\
{\normalsize School of Physics, Universiti Sains Malaysia}\\
{\normalsize 11800 USM Penang, Malaysia}}

\date{March 20, 2015}
\maketitle

\begin{abstract}
We report on the presence of new axially symmetric monopoles, antimonopoles and vortex-rings solutions of the SU(2)$\times$U(1) Weinberg-Salam model of electromagnetic and weak interactions. When the $\phi$-winding number $n=1$, and 2, the configurations are monopole-antimonopole pair (MAP) and monopole-antimonopole chain (MAC) with poles of alternating sign magnetic charge arranged along the $z$-axis. Vortex-rings start to appear from the MAP and MAC configurations when the winding number $n=3$. The MAP configurations possess zero net magnetic charge whereas the MAC configurations possess net magnetic charge of $4\pi n/e$. 

In the MAP configurations, the monopole-antimonopole pair is bounded by the ${\cal Z}^0$ field flux string and there is an electromagnetic current loop encircling it. The monopole and antimonopole possess magnetic charges $\pm\frac{4\pi n}{e}\sin^2\theta_W$ respectively. In the MAC configurations there is no string connecting the monopole and the adjacent antimonopole and they possess magnetic charges $\pm\frac{4\pi n}{e}$ respectively. The MAC configurations possess infinite total energy and zero magnetic dipole moment whereas the MAP configurations which are actually sphalerons possess finite total energy and magnetic dipole moment. The configurations were investigated for varying values of Higgs self-coupling constant $0\leq \lambda\leq 40$ at Weinberg angle $\theta_W=\frac{\pi}{4}$.
\end{abstract}

\section{Introduction}
Magnetic monopole was first introduced into the Maxwell theory by P.A.M. Dirac \cite{kn:1}. The presence of the magnetic monopole with pole strength $g$ leads to the requirement that all electric charges have to be quantized in integral multiples of a unit electric charge $e$ given by the formula $\frac{ge}{\hbar c} = \frac{1}{2}n$. The fact that electric charges are quantized and that there are no other explanation for this quantization makes magnet monopole a very important particle that has yet to be discovered. The magnetic field of the Dirac monopole carries a string singularity.

In 1969, a non-Abelian magnetic monopole with only a point singularity was found as a solution to the pure SU(2) Yang-Mills theory by Wu and Yang \cite{kn:2}. However both the Dirac and Wu-Yang monopole possess infinite energy due to the presence of a point singularity in the solution. It was in 1974 that a finite energy magnetic monopole was found by 't Hooft and Polyakov \cite{kn:3} independently in the SU(2) Georgi-Glashow model. The mass of the 't Hooft-Polyakov monopole was calculated to be of order 137 $M_W$, where $M_W$ is the mass of the intermediate vector boson. In the Georgi-Glashow model, $M_W<53$ GeV, however the mass is given by $M_W = 80.385 \pm 0.015$ GeV in the Particle Physics Booklet \cite{kn:4}. Hence the mass of the magnetic monopole in the Georgi-Glashow model is of the order of 11 TeV. 

A few years later in 1977, Y. Nambu found string-like configurations in the SU(2)$\times$U(1) Weinberg-Salam model \cite{kn:5}. These configurations are a monopole-antimonopole pair bound by a flux string of the ${\cal Z}^0$ field. The total energy of this MAP configuration is finite and the mass of the monopole and antimonopole together with the string is estimated to be in the TeV range. At asymptotically large distances, the real electromagnetic field is a linear combination of U(1) and SU(2) gauge fields created by the MAP.  Although the arguments and calculations presented are not rigorous, but the existence of massive string-like MAP configurations of the Weinberg-Salam theory had been accurately predicted by Nambu. Our numerical results for the 1-MAP and 2-MAP configurations given in Section \ref{sect.4.3} here confirmed Nambu's finding years ago \cite{kn:5}.

After Nambu's work \cite{kn:5}, there is a large amount of work done on the classical solutions of the Weinberg-Salam theory \cite{kn:5} - \cite{kn:13}, which is a hybrid of of the Abelian Maxwell and the non-Abelian Georgi-Glashow theory. A well known solution of the Weinberg-Salam theory is the ``sphaleron", first coined by Klinkhamer and Manton \cite{kn:7}, which possesses baryon number $Q_B\mbox{(sphaleron)}=\frac{1}{2}$. A sphaleron is particle-like, static, localized in space and unstable. In this solution, Klinkhamer and Manton \cite{kn:7} noticed that there is an electric current in the U(1) field. 
Other interesting work done on sphaleron include that of Hindmarsh and James \cite{kn:8} and Radu and Volkov \cite{kn:9} which state that within the sphaleron there is a monopole-antimonopole pair and a loop of electromagnetic current. 

Other work on sphalerons include the work of Ref. \cite{kn:10} and \cite{kn:11} which is a series of work using the same ansatz and definition. In these papers, no monopole-antimonopole pair or current loop is found in the sphaleron. The magnetic ansatz is used for the SU(2) gauge field which is almost the same ansatz as our work here but the solutions reported in Ref. \cite{kn:10} and \cite{kn:11} differ from our solutions. This is because the profile functions for the Higgs field differ from our profile functions of the Higgs field. Also the boundary conditions for their profile functions are different from our boundary conditions at small distances and along the $z$-axis. Their ansatz possesses $\theta$-winding number $m=1, 2, 3, 4, 5, 6$, whereas our ansatz possesses $\theta$-winding number $m=1$ only. Their numerical results did not reveal the inner structure of the sphaleron and anti-sphaleron. Their ansatz also by pass the MAC configurations. 

In 1997, Cho and Maison \cite{kn:12} reported on the single monopole configuration in the Weinberg-Salam theory which is the same as the first solution in our MAC sequence of solutions. Similar to the Georgi-Glashow model, this configuration is spherically symmetrical. This electrically charged single monopole possesses magnetic charge $4\pi/e$. The apparent string singularity of this monopole along the negative $z$-axis of the U(1) gauge field is a pure gauge artifact that can be removed with a hypercharge U(1) gauge transformation. Hence unlike the MAP solution \cite{kn:5} - \cite{kn:9}, this monopole does not possess a string. The total energy of this single monopole configuration is infinite due the point magnetic charge of the U(1) field. However by using various method discussed in Ref. \cite{kn:13}, this electroweak monopole mass is estimated to be about 4 to 10 TeV which is within the range of the recent MoEDAL detector at LHC, CERN \cite{kn:14}. Hence there is a possibility that this Cho-Maison monopole can be detected by the experiment.

In this paper, we present numerical MAP, MAC and vortex-rings configurations that are axially symmetrical. Similar to the Georgi-Glashow theory, the only spherically symmetrical monopole solution is the single monopole with magnetic charge $4\pi/e$ that was found by Cho and Maison \cite{kn:12}. The other monopole configurations are at most axially symmetrical. 

We solved the SU(2)$\times$U(1) Weinberg-Salam equations of motion numerically over all space for presence of new axially symmetric electrically neutral monopole configurations. The solutions found are monopole-antimonopole pairs (MAP) and monopole-antimonopole chains (MAC) configurations when the $\phi$-winding number $n=1$ and 2. Vortex-ring configurations start to appear from the MAP and MAC configurations when the winding number $n=3$. 

Our 1-MAP and 2-MAP configurations once again confirmed the findings of Ref. \cite{kn:6} - \cite{kn:9}. The MAP configurations possess zero net magnetic charge. The 1-MAP is a sphaleron with baryon number $Q_B=\frac{n}{2}$ and the 2-MAP is a sphaleron anti-sphaleron pair with baryon number $Q_B=0$. When the $\phi$-winding number $n=1$ and 2, the monopole-antimonopole pair is bounded by the ${\cal Z}^0$ field flux string. When $n=1$ and Weinberg angle $\theta_W=\frac{\pi}{4}$, the monopole and antimonopole possess magnetic charges $\pm\frac{2\pi}{e}$ respectively and hence they are half Cho-Maison monopole and antimonopole. 

Similar to the results discussed by others \cite{kn:5} - \cite{kn:9}, the MAP configurations or sphalerons found here possess finite total energy and magnetic dipole moment. Each monopole-antimonopole pair ($n=1, 2$) and vortex-ring ($n=3$) are surrounded by an electromagnetic current loop. The total energy is finite because there is no magnetic monopole presents in the U(1) field but just external electric current loops that provide the magnetic dipole moment of the U(1) field.

In the MAC configurations the monopoles and the antimonopoles are not held by the neutral ${\cal Z}^0$ flux. When $n=1$, the monopole and antimonopole possess magnetic charges $\pm\frac{4\pi}{e}$ respectively. Hence the monopole and antimonopole are whole Cho-Maison monopole and antimonopole. Since the MAC configurations possess odd number of poles, their net magnetic charge is $4\pi n/e$. The total energy of these MAC solutions is infinite due the point magnetic charge of the monopole in the U(1) field as discussed by Cho and Maison \cite{kn:12} for their one monopole solution. However the SU(2) part of the total energy is finite. This is expected as the energy of all the monopole solutions in the SU(2) Georgi-Glashow model is finite whereas the energy of a point charge in the Abelian U(1) theory blows up at the point. The MAC configurations do not possess magnetic dipole moment and unlike the MAP configurations, the magnetic charges of the poles do not change with Weinberg angle, $0\leq \theta_W \leq \frac{\pi}{2}$. 
 
In the next Section, we briefly present the Weinberg-Salam model and in Section \ref{sect.3} we obtained the reduced equations of motion by using the axially symmetrical magnetic ansatz. The MAC, MAP, and vortex-rings configurations are discussed and investigated in Section \ref{sect.4} for values of Higgs self-coupling constant $0\leq \lambda \leq 40$ at Weinberg angle $\theta_W=\frac{\pi}{4} $. When $n=1$ and 2, the MAC configurations presented are the one pole, three poles, and five poles solutions and the MAP configurations presented are the two poles and four poles solutions. When $n=3$, vortex-rings configurations are found. We end with some comments in Section \ref{sect.5}.

\section{The Standard Weinberg-Salam Model}
\label{sect.2}

The Lagrangian in the standard Weinberg-Salam model is given by \cite{kn:12}, \cite{kn:13}
\begin{eqnarray}
&&{\cal L} = -{(\cal D}_\mu \boldsymbol{\phi})^\dagger ({\cal D}^\mu \boldsymbol{\phi}) - \frac{\lambda}{2}\left(\boldsymbol{\phi}^\dagger \boldsymbol{\phi} -\nu^2\right)^2 - \frac{1}{4}{\bf F}_{\mu\nu}\cdot {\bf F}^{\mu\nu} - \frac{1}{4}f_{\mu\nu}f^{\mu\nu},
\label{eq.1}\\
&&{\cal D}_\mu \boldsymbol{\phi} = \left(D_\mu - \frac{ig^\prime}{2} a_\mu \right) \boldsymbol{\phi}, ~~D_\mu = \partial_\mu - \frac{ig}{2} \boldsymbol{\sigma} \cdot {\bf A}_\mu,
\label{eq.2}
\end{eqnarray}
where ${\cal D}_\mu$ is the covariant derivative of the SU(2)$\times$U(1) group and $D_\mu$ is the covariant derivative of the SU(2) group only. The gauge coupling constant, potentials, and electromagnetic fields of the SU(2) group are given by $g$, ${\bf A}_\mu = A^a_\mu (\frac{\sigma^a}{2i})$, and ${\bf F}_{\mu\nu} = F^a_{\mu\nu} (\frac{\sigma^a}{2i})$ respectively, whereas the U(1) group's gauge coupling constant, potentials, and electromagnetic fields are given $g^\prime$, $a_\mu$, and $f_{\mu\nu}$ respectively. The $\sigma^a$ are Pauli matrices. The complex scalar Higgs doublet is $\phi$, the Higgs field self-coupling constant is $\lambda$ and the mass of the Higgs boson $M_H=\nu\sqrt{2\lambda}$, where $\nu$ is the Higgs field vacuum expectation value. The masses of the ${\cal W}$ and ${\cal Z}$ bosons are given by $M_W=\nu\frac{g}{\sqrt{2}}$ and $M_Z=\nu\sqrt{\frac{g^2+g^{\prime 2}}{2}}$ respectively. The metric used is $-g_{00}=g_{11}=g_{22}=g_{33}=1$.

The equations of motion that follow from Lagrangian (\ref{eq.1}) are
\begin{eqnarray}
&&{\cal D^\mu}{\cal D_\mu}\boldsymbol{\phi} = \lambda\left(\boldsymbol{\phi}^\dagger\boldsymbol{\phi}-\nu^2\right)\boldsymbol{\phi},
\label{eq.3}\\
&&D^\mu {\bf F}_{\mu\nu} = -{\bf j}_\nu = \frac{ig}{2}\{\boldsymbol{\phi}^\dagger\boldsymbol{\sigma}({\cal D_\nu}\boldsymbol{\phi})-({\cal D_\nu}\boldsymbol{\phi})^\dagger\boldsymbol{\sigma\phi}\},
\label{eq.4}\\
&&\partial^\mu f_{\mu\nu} = -k_\nu = \frac{ig^\prime}{2}\{\boldsymbol{\phi}^\dagger({\cal D_\nu}\boldsymbol{\phi})-({\cal D_\nu}\boldsymbol{\phi})^\dagger\boldsymbol{\phi}\}.
\label{eq.5}
\end{eqnarray}
The Higgs field can also be written as 
\begin{eqnarray}
&&\boldsymbol{\phi} = \frac{{\cal H}(r,\theta)}{\sqrt{2}}\boldsymbol{\xi}, ~~~\left(\boldsymbol{\phi}^\dagger \boldsymbol{\phi} = \frac{{\cal H}^2}{2}, ~~\boldsymbol{\xi}^\dagger\boldsymbol{\xi}=1\right),\nonumber\\
&&\hat{\Phi}^a = \boldsymbol{\xi}^\dagger\sigma^a \boldsymbol{\xi}, ~~~\sigma^a = \left(\begin{array}{ll}                   
																																												\delta^a_3 & \delta^a_1-i\delta^a_2\\
																																					 \delta^a_1+i\delta^a_2 & -\delta^a_3
																																					               \end{array}\right)
\label{eq.6}
\end{eqnarray}
where $\frac{{\cal H}(r,\theta)}{\sqrt{2}}$ is the Higgs modulus, $\boldsymbol{\xi}$ is a column 2-vector, and $\hat{\Phi}^a$ is the Higgs field unit vector. The energy density of Lagrangian (\ref{eq.1}) is given by

\begin{eqnarray}
{\cal E} &=& \frac{1}{4}F^a_{ij}F^a_{ij} + \frac{1}{2}F^a_{i0}F^a_{i0} + \frac{1}{4}f_{ij}f_{ij} + \frac{1}{2}f_{i0}f_{i0} + ({\cal D}_i \boldsymbol{\phi})^\dagger ({\cal D}_i \boldsymbol{\phi}) \nonumber\\
&+& ({\cal D}_0 \boldsymbol{\phi})^\dagger ({\cal D}_0 \boldsymbol{\phi}) + \frac{\lambda}{2}\left(\boldsymbol{\phi}^\dagger \boldsymbol{\phi} -\nu^2\right)^2.
\label{eq.7}
\end{eqnarray}

\section{The Axially Symmetric Magnetic Ansatz}
\label{sect.3}

\subsection{The Ansatz}
The axially symmetric magnetic ansatz for the SU(2) gauge field, Higgs field, and U(1) gauge field are respectively given by \cite{kn:15} , \cite{kn:16}
\begin{eqnarray}
gA_i^a &=&  - \frac{1}{r}\psi_1(r, \theta) \hat{n}^{a}_\phi\hat{\theta}_i + \frac{1}{r}\psi_2(r, \theta)\hat{n}^{a}_\theta\hat{\phi}_i
+ \frac{1}{r}R_1(r, \theta)\hat{n}^{a}_\phi\hat{r}_i - \frac{1}{r}R_2(r, \theta)\hat{n}^{a}_r\hat{\phi}_i, \nonumber\\
gA^a_0 &=& \tau_r(r, \theta)\hat{n}^{a}_r + \tau_\theta(r, \theta)\hat{n}^{a}_\theta = A_0(r, \theta)\hat{g}^a,
\label{eq.8}\\
\Phi^a &=& \Phi_1(r, \theta)\hat{n}^a_r + \Phi_2(r, \theta)\hat{n}^a_\theta = \frac{{\cal H}(r, \theta)}{\sqrt{2}} \hat{\Phi}^a,\nonumber\\
\boldsymbol{\xi} &=& i\left(\begin{array}{l}                   
																																												\sin\frac{\alpha(r,\theta)}{2}e^{-in\phi}\\
																																					 							 -\cos\frac{\alpha(r,\theta)}{2}
																																					               \end{array}\right)\nonumber\\
\hat{\Phi}^a &=& \boldsymbol{\xi}^\dagger\sigma^a \boldsymbol{\xi} = -\hat{h}^a. 																																				      
\label{eq.9}\\
g^\prime B_\mu &=& B_0\delta^0_\mu + \frac{1}{r}B_1(r,\theta)\hat{\phi}_i\delta^i_\mu.
\label{eq.10}
\end{eqnarray}

In the rectangular coordinate system, the unit vectors, \cite{kn:17}
\begin{eqnarray}
\hat{h}^a&=&h_1(r,\theta)\hat{n}^{a}_r + h_2(r,\theta)\hat{n}^{a}_\theta =  \cos(\alpha-\theta)\hat{n}^{a}_r + \sin(\alpha-\theta)\hat{n}^{a}_\theta,\nonumber\\
&=&\sin\alpha \cos n\phi ~\delta^{a1} + \sin\alpha \sin n\phi ~\delta^{a2} + \cos\alpha ~\delta^{a3},
\label{eq.11}\\
\nonumber\\
\hat{g}^a&=&g_1(r,\theta)\hat{n}^{a}_r + g_2(r,\theta)\hat{n}^{a}_\theta =  \cos(\gamma-\theta)\hat{n}^{a}_r + \sin(\gamma-\theta)\hat{n}^{a}_\theta,\nonumber\\
 &=&\sin\gamma \cos n\phi ~\delta^{a1} + \sin\gamma \sin n\phi ~\delta^{a2} + \cos\gamma ~\delta^{a3}, ~~~\gamma=\gamma(r,\theta),
\label{eq.12}
\end{eqnarray}
where $\cos\alpha = \left\{h_1\cos \theta - h_2\sin \theta\right\}$ and $\sin\alpha = \left\{h_1\sin \theta + h_2\cos \theta\right\}$.
The profile functions of the time component SU(2) gauge potential can be written as $\tau_r=A_0 g_1$ and $\tau_\theta=A_0 g_2$, and $g_1\rightarrow h_1$, $g_2\rightarrow h_2$ at asymptotically large $r$. 

From our previous work in the SU(2) Georgi-Glashow \cite{kn:19}, we know that for the MAC, MAP, and vortex-rings solutions, the angle $\alpha(r, \theta) \rightarrow p\,\theta$ as $r\rightarrow \infty$, where $p=1, 2, 3, ...$, is a natural number representing the number of magnetic poles (monopoles and antimonopoles) in the configuration. When $p$ is odd, we get the MAC solutions and when $p$ is even, we get the MAP solutions.
The spatial spherical coordinate unit vectors are
$\hat{r}_i = \sin\theta ~\cos \phi ~\delta_{i1} + \sin\theta ~\sin \phi ~\delta_{i2} + \cos\theta~\delta_{i3}$, 
$\hat{\theta}_i = \cos\theta ~\cos \phi ~\delta_{i1} + \cos\theta ~\sin \phi ~\delta_{i2} - \sin\theta ~\delta_{i3}$, 
$\hat{\phi}_i = -\sin \phi ~\delta_{i1} + \cos \phi ~\delta_{i2}$, whereas the isospin coordinate unit vectors with $\phi$-winding number $n=1, 2, 3, ...$ are given by
\begin{eqnarray}
&&\hat{n}_r^a = \sin \theta ~\cos n\phi ~\delta_{1}^a + \sin \theta ~\sin n\phi ~\delta_{2}^a + \cos \theta~\delta_{3}^a, \nonumber\\
&&\hat{n}_\theta^a = \cos \theta ~\cos n\phi ~\delta_{1}^a + \cos \theta ~\sin n\phi ~\delta_{2}^a - \sin \theta ~\delta_{3}^a, \nonumber\\
&&\hat{n}_\phi^a = -\sin n\phi ~\delta_{1}^a + \cos n\phi ~\delta_{2}^a.
\label{eq.13}
\end{eqnarray}

We also notice that \cite{kn:12}
\begin{eqnarray}
C_i = i\boldsymbol{\xi}^\dagger\partial_i \boldsymbol{\xi} = \frac{n(1-\cos\alpha)}{2r\sin\theta}\hat{\phi}_i
\label{eq.14}
\end{eqnarray}
and upon applying the Cho Abelian decomposition \cite{kn:17a} on the spatial SU(2) gauge potential of Eq. (\ref{eq.8}), we have \cite{kn:17b}
\begin{eqnarray}
A^a_i &=& \hat{A}^a_i + X^a_i, ~~~\hat{A}^a_i=A\hat{\phi}_i\hat{\Phi}^a-\frac{1}{g}\epsilon^{abc}\hat{\Phi}^b\partial_i\hat{\Phi}^c, ~~~gA=-\frac{1}{r}(\psi_2 h_2 - R_2 h_1),\nonumber\\
X^a_i &=& X_1\hat{\phi}_i \hat{\Phi}^a_1 + (X_3\hat{r}_i + X_4\hat{\theta}_i)\hat{\phi}^a, ~~~\hat{\Phi}^a_1 = h_2\hat{n}^a_r - h_1\hat{n}^a_\theta,
\label{eq.15}\\
gX_1 &=& -\frac{1}{r}(\psi_2 h_1 + R_2 h_2) + \frac{n\sin\alpha}{r\sin\theta}, ~gX_3 = \frac{1}{r}(r\alpha^\prime + R_1), ~gX_4 = \frac{1}{r}(\dot{\alpha}-\psi_1). \nonumber
\end{eqnarray}
The electromagnetic field strength tensor of the Cho decomposed gauge potential $A^a_i$ (\ref{eq.15}) is given by \cite{kn:17a}, \cite{kn:17b}
\begin{eqnarray}
F^a_{ij} = \hat{F}^a_{ij} + \hat{D}_iX^a_j - \hat{D}_jX^a_i + g\epsilon^{abc}X^b_i X^c_j
\label{eq.16}
\end{eqnarray}
where $\hat{F}^a_{ij}=\hat{F}_{ij}\hat{\Phi}^a$ is the field strength of the self-dual potential $\hat{A}^a_i$ and the covariant derivative, $\hat{D}_i\hat{\Phi}^a=\partial_i\hat{\Phi}^a + g\epsilon^{abc}\hat{A}^b\hat{\Phi}^c=0$, vanishes. The Abelian electromagnetic tensor
\begin{eqnarray}
\hat{F}_{\mu\nu} &=& \partial_\mu A_\nu - \partial_\nu A_\mu + \frac{1}{g}\epsilon^{abc}\hat{\Phi}^a\partial_\mu\hat{\Phi}^b\partial_\nu\hat{\Phi}^c, 
\label{eq.17}\\
gA_\mu &=& gA\hat{\phi}_i\delta^i_\mu - (g_1 h_1 + g_2 h_2)\tau \delta^0_\mu,\nonumber
\end{eqnarray}
is the 't Hooft's electromagnetic field upon symmetry breaking and a suitable Abelian gauge potential is given by
\begin{eqnarray}
g{\cal A}^H_\mu &=& gA_\mu + 2 C_\mu = -\frac{1}{r}A_1\hat{\phi}_i\delta^i_\mu - (g_1 h_1 + g_2 h_2)\tau \delta^0_\mu,\nonumber\\
A_1 &=& \left\{(\psi_2h_2 - R_2h_1) - \frac{n(1-\cos\alpha)}{\sin\theta}\right\}
\label{eq.18}
\end{eqnarray}

\subsection{The Equations of Motion}
When the magnetic ansatz (\ref{eq.8}), (\ref{eq.9}) and (\ref{eq.10}) is substituted into the equations of motion (\ref{eq.3}) to (\ref{eq.5}), the equations of motion (\ref{eq.3}) reduced to the two following partial second order coupled nonlinear equations,

\begin{eqnarray}
&&\partial^i\partial_i{\cal H} - \frac{\lambda}{2}\left({\cal H}^2-2\nu^2\right){\cal H} - \frac{1}{4r^2}\left\{\left(\psi_1-\left[1-\frac{\dot{h}_1}{h_2}\right]\right)^2 + \left(R_1-\frac{rh_1^\prime}{h_2}\right)^2\right\}{\cal H}\nonumber\\
&& + \frac{1}{4r^2}\left\{(A_1+n\csc\theta)^2 -(A_1-B_1)^2 - (n-\psi_2)^2 - (R_2-n\cot\theta)^2\right\}{\cal H}\nonumber\\
&& +\frac{1}{4}\left\{A_0^2(g_1h_2-g_2h_1)^2 + (B_0-A_0[g_1h_1+g_2h_2])^2\right\}{\cal H}=0,
\label{eq.19}\\\nonumber\\
&&\cot\theta - r^2\left\{\frac{\partial^i\partial_ih_1}{h_2}-\frac{\partial^ih_1\partial_ih_2}{h_2^2}\right\} - \frac{1}{\sin\theta}(\dot{\psi_1\sin\theta}) + (rR_1)^\prime\nonumber\\
&&+ 2r \left\{R_1-\frac{rh_1^\prime}{h_2}\right\}(\ln{\cal H})^\prime - 2\left\{\psi_1-\left[1-\frac{\dot{h}_1}{h_2}\right]\right\}\dot{(\ln{\cal H})} \nonumber\\
&&+ (B_1+n\csc\theta)\left(h_1\psi_2+h_2R_2 - n(h_1+h_2\cot\theta)\right) + r^2 B_0 A_0(g_1h_2-g_2h_1) = 0,\nonumber\\
\mbox{or} && \frac{1}{2r^2}{\cal H}=0.
\label{eq.20}
\end{eqnarray}

\noindent Here ``prime" and ``dot'' mean $\frac{\partial}{\partial r}$ and $\frac{\partial}{\partial \theta}$ respectively. 
The equations of motion (\ref{eq.4}) reduced to the six following partial second order coupled nonlinear equations,

\begin{eqnarray}
 j^a_j&=&\partial^iF^a_{ij}+\epsilon^{abc}gA^{bi}F^c_{ij}+\epsilon^{abc}gA^{b0}F^c_{0j}\nonumber\\
&=& \frac{g}{4r}{\cal H}^2\left\{(\psi_2 h_1+R_2 h_2 - n[h_1+h_2\cot\theta])\hat{h}^a_\bot\hat{\phi}_j + (A_1-B_1)\hat{h}^a \hat{\phi}_j\right\}\nonumber\\
&+& \frac{g}{4r}{\cal H}^2\left\{\left(R_1-\frac{rh_1^\prime}{h_2}\right)\hat{n}^a_\phi \hat{r}_j - \left(\psi_1-\left[1-\frac{\dot{h}_1}{h_2}\right]\right)\hat{n}^a_\phi \hat{\theta}_j\right\},
\label{eq.21}\\\nonumber\\
j^a_0&=&\partial^iF^a_{i0}+\epsilon^{abc}gA^{bi}F^c_{i0}\nonumber\\
&=&\frac{g}{4}{\cal H}^2\left\{(A_0[g_1 h_1 + g_2 h_2]	-  B_0)\hat{h}^a + A_0(g_2 h_1 - g_1 h_2)\hat{h}^a_\bot\right\},	
\label{eq.22}
\end{eqnarray}
where the unit vector $\hat{h}^a_\bot = -h_2\hat{n}^a_r + h_1\hat{n}^a_\theta$ is perpendicular to $\hat{h}^a$. The SU(2) electric current source density is $j^a_i$ and the SU(2) electric charge source density is $j^a_0$.
The equations of motion (\ref{eq.5}) reduced to the two following partial second order coupled nonlinear equations,

\begin{eqnarray}
g^\prime k_j &=& \left\{\partial^i\partial_i\left(\frac{1}{r}B_1\right) - \frac{1}{r^3\sin^2\theta}B_1\right\}\hat{\phi}j = \frac{g^{\prime 2}}{4r}{\cal H}^2(B_1-A_1)\hat{\phi}_j,
\label{eq.23}\\
\nonumber\\
g^\prime k_0 &=& \partial^i\partial_iB_0  = \frac{g^{\prime 2}}{4}{\cal H}^2\{B_0 - A_0(g_1 h_1 + g_2 h_2)\},
\label{eq.24}											 
\end{eqnarray}
where $k_i$ is the U(1) electric current source density and $k_0$ is the U(1) electric charge source density.

There are all together ten reduced electrically charged equations of motion (\ref{eq.14})-(\ref{eq.19}). In the work here, we solved for the electrically neutral monopole configurations by setting $A_0$ and $B_0$ to zero and the total number of equations of motion is reduced to only seven equations. In this case, both the electric charge source density $j^a_0$ and $k_0$ vanish. However the electric current source density $j^a_i$ and $k_i$ do not necessarily vanish. They however vanish at large $r$ in the Higgs vacuum.

\subsection{The Energy}
In the electrically neutral monopole configuration, the energy density (\ref{eq.7}) can be written as 
\begin{eqnarray}
{\cal E} &=&  \frac{1}{4g^2}(gF^a_{ij})(gF^a_{ij}) + \frac{1}{4g^{\prime 2}}(g^{\prime}f_{ij})(g^{\prime}f_{ij})\nonumber\\
 &+& \frac{1}{2}\partial^i{\cal H}\partial_i{\cal H} + \frac{1}{2}{\cal H}^2({\cal D}^i\xi)^\dagger({\cal D}_i\xi) + \frac{\lambda}{8}\left({\cal H}^2 -2\nu^2\right)^2,
\label{eq.25}
\end{eqnarray}
where
\begin{eqnarray}
({\cal D}^i\xi)^\dagger({\cal D}_i\xi)&=&\frac{1}{4}\partial^i\alpha\,\partial_i\alpha + \frac{n^2(1-\cos\alpha)}{2r^2\sin^2\theta} + \frac{n}{2}(1-\cos\alpha)(g^\prime B^i)\partial_i\phi\nonumber\\
&+&\frac{1}{2}\{\hat{n}_\phi^a \partial^i\alpha + n\,\partial^i\phi\,[\hat{n}_r^a\cos\theta - \hat{n}_\theta^a\sin\theta - \hat{h}^a]\}(gA^a_i)\nonumber\\
&+&\frac{1}{4}(gA^{ai})(gA^a_i)-\frac{1}{2}(g^\prime B^i)(gA^a_i)\hat{h}^a + \frac{1}{4}(g^\prime B^i)(g^\prime B_i).
\label{eq.26}
\end{eqnarray}
The total energy of the MAC configurations is infinite due to $(g^{\prime}f_{ij})(g^{\prime}f_{ij})$ which is singular at the location of the monopoles. However the energy density (\ref{eq.20}) is regular over all space for the MAP configurations. Hence the MAP total energy $E = \frac{e}{4\pi}\int{{\cal E}_n}\,d^3x$ is finite.

\subsection{The Unitary Gauge} 

In order to determine the electric and magnetic charge of the electromagnetic weak monopole configuration \cite{kn:12}, the gauge potentials $A^a_\mu$ and Higgs field $\Phi^a$ of Eq. (\ref{eq.8}) are gauge transformed to $A^{\prime a}_\mu$ and $\Phi^{\prime a}=\delta^a_3$ in the unitary gauge. Using the gauge transformation,

\begin{eqnarray}
&&U_1 = -i\left[\begin{array}{ll}                   
																 \cos\frac{\alpha}{2}  				  & \sin\frac{\alpha}{2} e^{-in\phi}\\
																 \sin\frac{\alpha}{2} e^{in\phi} & -\cos\frac{\alpha}{2}
																 \end{array}\right]
= \cos\frac{\Theta_1}{2} + i\hat{u}_r^a \sigma^a \sin\frac{\Theta_1}{2}, 														 
\label{eq.27}\\
&&\Theta_1=-\pi ~~\mbox{and}~ ~\hat{u}_r^a = \sin\frac{\alpha}{2}\cos n\phi \delta^a_1 + \sin\frac{\alpha}{2}\sin n\phi \delta^a_2 + \cos\frac{\alpha}{2} \delta^a_3,	\nonumber	
\end{eqnarray}																
																 
\noindent we obtain the transformed Higgs column unit vector and the SU(2) gauge potentials which are respectively given by 

\begin{eqnarray}
\xi^\prime &=& U\xi = \left[\begin{array}{l}                   
																 0\\
																 1
																 \end{array}\right]\nonumber\\
gA^{\prime a}_\mu &=& -gA^a_\mu - \frac{2}{r}\left\{\psi_2\sin\left(\theta-\frac{\alpha}{2}\right) + R_2\cos\left(\theta-\frac{\alpha}{2}\right)\right\}\hat{u}_r^a\,\hat{\phi}_\mu\nonumber\\
 &-& \partial_\mu\alpha\,\hat{u}_\phi^a - \frac{2n\sin\frac{\alpha}{2}}{r\sin\theta}\hat{u}_\theta^a\,\hat{\phi}_\mu															 
\label{eq.28}\\
&+& 2\left\{\tau_r\cos\left(\theta-\frac{\alpha}{2}\right) - \tau_\theta\sin\left(\theta-\frac{\alpha}{2}\right)\right\}\hat{u}_r^a\,\delta^0_\mu,\nonumber
\end{eqnarray}																
or

\begin{eqnarray}
gA^{\prime 1}_\mu &=& -\frac{\cos n\phi}{r} \left\{\psi_2 h_1 + R_2 h_2 - \frac{n\sin\alpha}{\sin\theta}\right\}\hat{\phi}_\mu 
- \frac{\sin n\phi}{r}\left\{\,(\psi_1-\partial_\theta \alpha)\,\hat{\theta}_\mu - (R_1+r\partial_r \alpha)\,\hat{r}_\mu\right\}	\nonumber\\
&+& \left\{\tau_r\sin(\alpha-\theta) - \tau_\theta\cos(\alpha-\theta)\right\}\cos n\phi\,\delta^0_\mu\nonumber\\
&=& g\cos n\phi X_1\hat{\phi}_\mu + g\sin n\phi\{X_4\hat{\theta}_\mu + X_3\hat{r}_\mu\} + (g_1 h_2 - g_2 h_1)\tau \cos n\phi\,\delta^0_\mu
\label{eq.29}\\
gA^{\prime 2}_\mu &=& -\frac{\sin n\phi}{r} \left\{\psi_2 h_1 + R_2 h_2 - \frac{n\sin\alpha}{\sin\theta}\right\}\hat{\phi}_\mu 
+ \frac{\cos n\phi}{r}\left\{(\psi_1-\partial_\theta \alpha)\,\hat{\theta}_\mu - (R_1+r\partial_r \alpha)\,\hat{r}_\mu\right\}	\nonumber\\							 
&+& \left\{\tau_r\sin(\alpha-\theta) - \tau_\theta\cos(\alpha-\theta)\right\}\sin n\phi\,\delta^0_\mu\nonumber\\
&=& g\sin n\phi X_1\hat{\phi}_\mu - g\cos n\phi \{X_4\hat{\theta}_\mu + X_3\hat{r}_\mu\}  + (g_1 h_2 - g_2 h_1)\tau \sin n\phi\,\delta^0_\mu
\label{eq.30}\\
gA^{\prime 3}_\mu &=& \frac{1}{r}\left\{\psi_2 h_2 - R_2 h_1 - \frac{n(1-\cos\alpha)}{\sin\theta}\right\}\hat{\phi}_\mu
+ \left\{\tau_r\cos(\alpha-\theta) + \tau_\theta\sin(\alpha-\theta)\right\}\delta^0_\mu\nonumber\\
&=& \frac{1}{r} A_1 \hat{\phi}_\mu + (g_1 h_1 + g_2 h_2)\tau \delta^0_\mu
\label{eq.31}
\end{eqnarray}
Here we note that the gauge potential $gA^{\prime 3}_\mu$ (\ref{eq.31}) is actually the negative gauge potential of the 't Hooft electromagnetic field strength \cite{kn:3}, $\hat{F}_{\mu\nu} =\hat{\Phi}^a F^a_{\mu\nu} - \frac{1}{g}\epsilon^{abc}\hat{\Phi}^{a}D_{\mu}\hat{\Phi}^{b}D_{\nu}\hat{\Phi}^c = \partial_{\mu}A_\nu - \partial_{\nu}A_\mu - \frac{1}{g}\epsilon^{abc}\hat{\Phi}^{a}\partial_{\mu}\hat{\Phi}^{b}\partial_{\nu}\hat{\Phi}^c$, where $A_\mu = \hat{\Phi}^{a}A^a_\mu$ and $\hat{\Phi}^a = \Phi^a/|\Phi|$ as $\hat{\Phi}^{\prime a}=\delta^a_3$. Hence the gauge potential $gA^{\prime 3}_\mu = -g{\cal A}^H_\mu$ .

\subsection{The Electromagnetic and Neutral Fields}
The electromagnetic potential ${\cal A}_\mu$ and the neutral potential ${\cal Z}_\mu$ are defined as

\begin{eqnarray}
\left[\begin{array}{l}                   
			{\cal A}_\mu\\
			{\cal Z}_\mu
			\end{array}\right] = \left[\begin{array}{ll} 
			                      \cos\theta_W & \sin\theta_W \\
			                     -\sin\theta_W &  \cos\theta_W 
			                      \end{array}\right]         
\left[\begin{array}{l} 
B_\mu \\
A^{\prime 3}_\mu 
\end{array}\right]\nonumber\\\nonumber\\
= \frac{1}{\sqrt{g^2+g^{\prime 2}}}	\left[\begin{array}{ll} 
			                             g & g^\prime \\
			                     -g^\prime &  g
			                      \end{array}\right]         
\left[\begin{array}{l} 
B_\mu \\
A^{\prime 3}_\mu 
\end{array}\right]	                     			                   															 
\label{eq.32}
\end{eqnarray}	

\noindent where $\cos\theta_W=\frac{g}{\sqrt{g^2+g^{\prime 2}}}$ and the electric charge $e=\frac{g g^\prime}{\sqrt{g^2+g^{\prime 2}}}$. Hence we can write the electromagnetic gauge potential and the neutral ${\cal Z}^0$ gauge potential as
\begin{eqnarray}
{\cal A}_\mu &=& \frac{1}{\sqrt{g^2+g^{\prime 2}}}(g B_\mu + g^\prime A^{\prime 3}_\mu) \nonumber\\
&=& \frac{1}{e} \left(\cos^2\theta_W g^\prime B_\mu + \sin^2\theta_W g A^{\prime 3}_\mu\right)
\label{eq.33}\\
{\cal Z}_\mu &=& \frac{1}{\sqrt{g^2+g^{\prime 2}}}(-g^\prime B_\mu + g A^{\prime 3}_\mu) \nonumber\\
&=& \frac{1}{e}\cos\theta_W \sin\theta_W\left(-g^\prime B_\mu + g A^{\prime 3}_\mu\right).
\label{eq.34}
\end{eqnarray}

\section{The Results}
\label{sect.4}

The Weinberg-Salam equations of motion (\ref{eq.19})-(\ref{eq.24}) are solved numerically using the Maple and MATLAB software for MAC, MAP, and vortex-ring configurations. The MAC configurations are obtained when the total number of poles along the $z$-axis are odd. The solutions discussed here are (i) single monopole (M), (ii) the three poles, monopole-antimonopole-monopole (MAM), and (v) the five poles (MAMAM) configurations.
The MAP configurations are obtained when the total number of poles along the $z$-axis are even. The solutions discussed here are (i) the monopole-antimonopole pair or 1-MAP (MA) and (ii) the two monopole-antimonopole pairs or 2-MAP (MAMA) configurations. When $n=3$, vortex-rings are formed from both the MAC and MAP configurations.

\subsection{Numerical Procedure}
\label{sect.4.1}

The profile functions of the time component of the U(1) and SU(2) gauge fields, $A_0$ and $B_0$ are set to zero for the electrically neutral solutions. Hence the equations of motion Eq. (\ref{eq.22}) and (\ref{eq.24}) vanish identically and we are left with seven reduced coupled second order partial differential equations of motion, Eq. (\ref{eq.19}) - (\ref{eq.21}) and Eq. (\ref{eq.23}), to solve. We solve the numerical monopole solutions here for all space by solving for the profiles functions, $\psi_1$, $\psi_2$, $R_1$, $R_2$, $\Phi_1$, $\Phi_2$, and $B_1$. The gauge coupling constant $g$ and the Higgs field vacuum expectation value $\nu$ are set to unity, that is $g=\nu=1$. The Higgs self-coupling constant $\lambda$ is varied from zero to 40, $0\leq \lambda\leq 40$. 
The seven reduced equations of motion are then solved by fixing boundary conditions at small distances ($r\rightarrow 0$), large distances ($r\rightarrow \infty$), and along the $z$-axis at $\theta=0$ and $\pi$.

The asymptotic solutions at large $r$ are the self-dual solutions of the SU(2) Georgi-Glashow theory \cite{kn:18}, \cite{kn:19}, which determines the type of monopole configuration in the SU(2) gauge field and the Maxwell gauge potential $g^\prime B_i|_{r\rightarrow\infty}= g A^{\prime 3}_i$,
\begin{eqnarray}
\psi_1 &=& \dot{\alpha}, ~~\psi_2 =n\left\{1+ \frac{\sin(\alpha-\theta)}{\sin\theta}(a\cos\theta+b)\right\}\nonumber\\
R_1 &=& 0, ~~R_2 = n\left\{\cot\theta - \frac{\cos(\alpha-\theta)}{\sin\theta}(a\cos\theta+b)\right\}\nonumber\\
\Phi_1 &=& g\zeta \cos(\alpha-\theta), ~~\Phi_2 = g\zeta \sin(\alpha-\theta), ~a, b = \mbox{constant},\nonumber\\
B_G &=& B_1  + \frac{n(1-\cos\alpha)}{\sin\theta} = A_1  + \frac{n(1-\cos\alpha)}{\sin\theta}\nonumber\\
&=& \frac{n(\{a+b\}-\cos\alpha)}{\sin\theta} - \frac{na(1-\cos\theta)}{\sin\theta}.
\label{eq.35}
\end{eqnarray}
The function $\alpha(r,\theta)|_{r\rightarrow\infty}=p\,\theta$ where $p$ is a natural number that determines the total number of magnetic monopoles and antimonopoles in the configurations. In the MAC configurations, the parameter $p$ is an odd integer, $a=1$ and $b=0$. In the MAP configurations, the parameter $p$ is an even integer, $a=0$ and $b=1$. The profile function $B_1$ vanishes as $r\rightarrow\infty$ only for the MAP solutions. For the MAC solutions, $B_1 = - \frac{n(1-\cos\theta)}{\sin\theta}$ as ${r\rightarrow\infty}$. Hence for the MAP configurations, we solve for the profile function $B_1$ whereas for the MAC solutions, we solve for the profile function $B_G$ instead as $B_1$ is singular along the negative $z$-axis as $r\rightarrow\infty$ . We solved the equations of motion (\ref{eq.19})-(\ref{eq.24}) when $p= 1,2,3,4,5$ and the $\phi$-winding number $n=1,2$, and 3 for MAC and MAP configurations. 

Since the U(1) gauge potential $g^\prime B_i$ of the monopole solutions presented here approaches the 't Hooft gauge potential $-g{\cal A}^H_i$  at large $r$, the neutral gauge potential ${\cal Z}_\mu$ (\ref{eq.34}) therefore vanishes as $r\rightarrow\infty$ and this neutral ${\cal Z}^0$ field carry zero net electric and net magnetic charge as expected. 
The electromagnetic gauge potential ${\cal A}_\mu \rightarrow \frac{1}{e}(g^\prime B_\mu) = \frac{1}{e}(-g{\cal A}^H_\mu)$ when $r$ goes to infinity. Since the boundary conditions for the MAC and MAP solutions are such that $B_1 \rightarrow -\frac{n(1-\cos\theta)}{\sin\theta}$ and $B_1 \rightarrow 0$ respectively at $r$ infinity, the MAP configurations possess zero net magnetic charge whereas the MAC configurations possess net magnetic charge $\frac{4\pi n}{e}$. 

The electromagnetic dipole moment $\mu_m$ of the MAP configurations can also be calculated by using the boundary condition at large $r$,
\begin{eqnarray}
{\cal A}_i \rightarrow \frac{1}{e}(g^\prime B_i) = \frac{1}{er}B_1\hat{\phi}_i = -\frac{\mu_m\sin\theta}{r^2}\hat{\phi}_i.
\label{eq.36}
\end{eqnarray}
Hence $r B_1=-e\mu_m\sin\theta$ and by plotting the numerical result for $r B_1$, we can read the magnetic dipole moment for the MAP solutions in unit of $\frac{1}{e}$ at $\theta=\frac{\pi}{2}$.

The asymptotic solutions at small $r$ are the trivial solution and for both MAP and MAC configurations,
\begin{eqnarray}
\psi_1(0, \theta) = \psi_2(0, \theta) = R_1(0, \theta) = R_2(0, \theta) &=& 0,\nonumber\\
\mbox{(MAP)} ~~B_1(0, \theta) = 0  ~~~\mbox{or}~~~ \mbox{(MAC)} ~~B_G(0, \theta) &=& 0,\nonumber\\
\sin\theta ~\Phi_1(0,\theta) + \cos\theta ~\Phi_2(0,\theta) &=& 0,\nonumber\\
\left.\frac{\partial}{\partial r}\left\{\cos\theta ~\Phi_1(r,\theta) - \sin\theta ~\Phi_2(r,\theta)\right\}\right|_{r=0} &=& 0,
\label{eq.37}
\end{eqnarray}

\noindent The boundary condition along the $z$-axis at $\theta=0$ and $\pi$ for the MAP and MAC configurations is
\begin{eqnarray}
\partial_\theta \psi_1 = \partial_\theta \psi_2 = R_1 = R_2 = \partial_\theta \Phi_1 = \Phi_2 = 0, \nonumber\\
\mbox{(MAP)} ~~B_1 = 0  ~~~\mbox{or}~~~ \mbox{(MAC)} ~~B_G = 0,
\label{eq.38}
\end{eqnarray}

The monopole solutions are solved numerically using the mathematical software, Maple and MATLAB, by fixing the boundary conditions (\ref{eq.35}), (\ref{eq.37}) and (\ref{eq.38}) when $r=0$, $r=\infty$, $\theta=0$, and $\theta=\pi$ \cite{kn:16}, \cite{kn:18}, \cite{kn:19}.  Using the finite difference approximation method, the seven reduced equations of motion (\ref{eq.19}) - (\ref{eq.24}) are converted into a system of nonlinear equations which is then discretized onto a non-equidistant grid of size $70\times 60$ (MAC) and $70\times 120$ (MAP) covering the integration regions $0\leq \bar{x} \leq 1$ and $0\leq \theta\leq \pi$. The compactified coordinate $\bar{x} = \frac{r}{r+1}$. Upon replacing the partial derivative ~$\partial_r \rightarrow (1-\bar{x})^2 \partial_{\bar{x}}$~ and ~$\frac{\partial^2}{\partial r^2} \rightarrow (1-\bar{x})^4\frac{\partial^2}{\partial \bar{x}^2} - 2(1-\bar{x})^3\frac{\partial}{\partial \bar{x}}$~, the Jacobian sparsity pattern of the system was constructed by using Maple. The system of nonlinear equations is then solved numerically by MATLAB using the constructed Jacobian sparsity pattern, the trust-region-reflective algorithm, and a good initial starting solution. The overall error in the numerical results is estimated at $10^{-4}$.

\subsection{MAC/Vortex-ring Configurations} 
\label{sect.4.2}

The MAC solutions obtained are the single pole $n$-M ($p = 1, n=1,2,3$), the three poles $n$-MAM ($p = 3, n=1,2$) , and the five poles $n$-MAMAM ($p = 5, n=1,2$) configurations. When the $\phi$-winding number $n=3$ and $p=3, 5$, two vortex-rings with centers along the $z$-axis and equidistance from the origin together with one monopole located at $r=0$ are formed. The profiles functions, $\psi_1$, $\psi_2$, $R_1$, $R_2$, $\Phi_1$, $\Phi_2$, and $B_G$ for the MAC solutions are obtained numerically and they are all bounded functions of $r$ and $\theta$.

\begin{figure}[tbh]
\centering
\hskip-0.1in
\includegraphics[width=5.5in,height=8.0in]{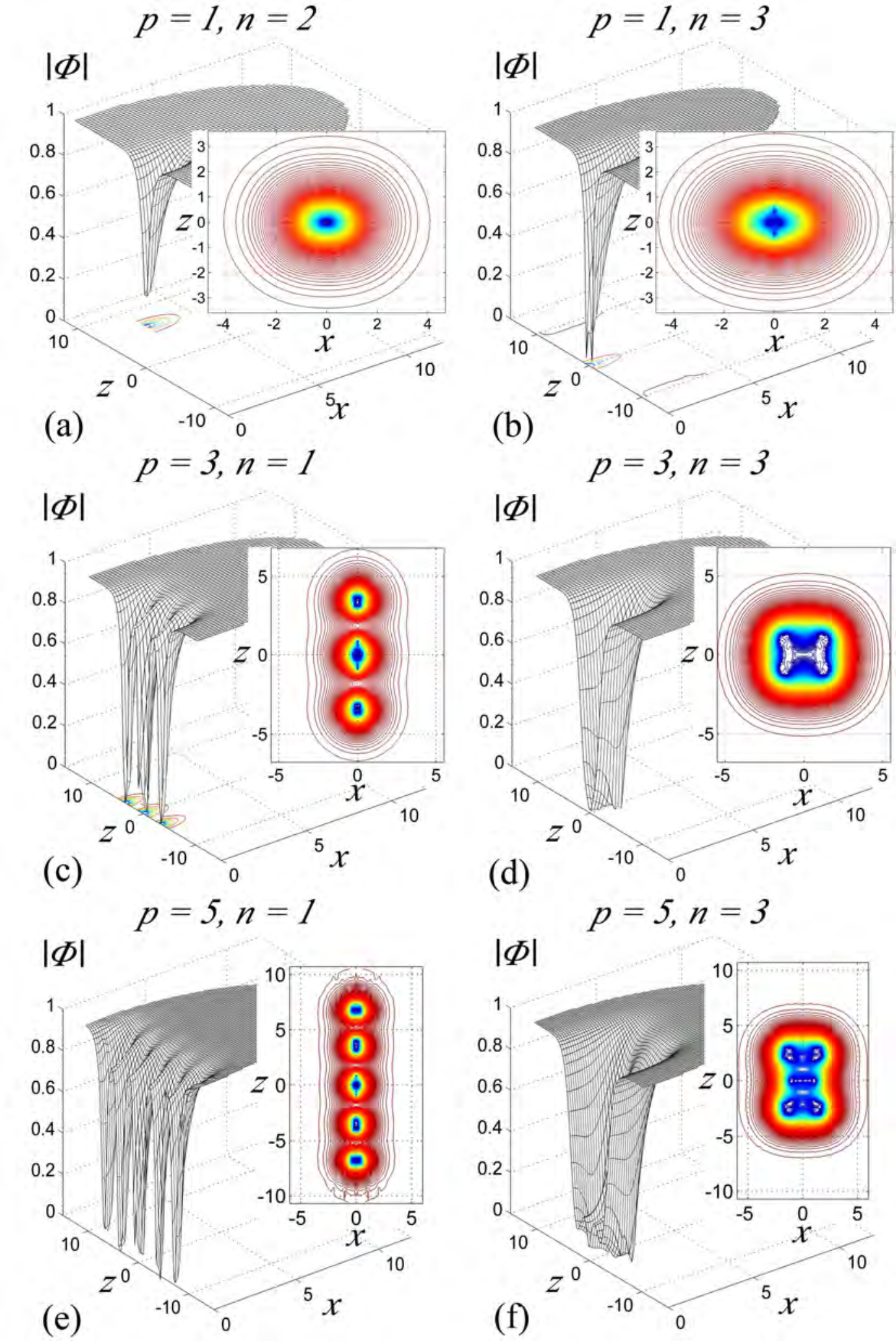} 
\caption{3D and contour line plot of the Higgs field modulus $|\Phi|$ along the $x$-$z$ plane for the one monopole, $n$-M solutions when (a) $n=2$ and (b) $n=3$, for the $p=3$, $n$-MAM solutions when (c) $n=1$ and (d) $n=3$, and for the $p=5$, $n$-MAMAM solutions when (e) $n=1$ and (f) $n=3$. Here $\lambda=\nu=1$ and $\theta_W=\frac{\pi}{4}$.}             
\label{fig.1}
\end{figure}

The Higgs field modulus $\Phi$ for the MAC configurations are shown in Figure \ref{fig.1} by 3D plot and contour line plot along the $x$-$z$ plane when the Higgs field self-coupling constant $\lambda=1$, the Higgs expectation value $\nu=1$ and the Weinberg angle $\theta_W=\frac{\pi}{4}$. Figure \ref{fig.1} (a) and (b) show the Higgs field modulus plots for the one monopole $n$-M solutions when $n=2$ and 3 respectively. Figure \ref{fig.1} (c) and (d) show the Higgs field modulus plots for the three poles $n$-MAM solutions when $n=1$ and 3 respectively and Figure \ref{fig.1} (e) and (f) show the Higgs field modulus plots for the five poles $n$-MAMAM solutions when $n=1$ and 3 respectively. The locations of the magnetic monopoles and vortex-rings are read from the zeros of the Higgs field modulus and tabulated in Table \ref{table.1} as $d_{(\rho,z)} = (\rho_i, \pm z_i)$ for $n=1$, 2, 3, $\lambda=\nu=1$ and $\theta_W=\frac{\pi}{4}$. Here $\pm z_i$ is the position of the $i$-th pole (or the center of the $i$-th vortex-ring along the $z$-axis) and $\rho_i$ is the radius of the $i$-th vortex-ring.
The Higgs field modulus for the MAC/vortex-ring configurations are almost similar to those of the SU(2) Georgi-Glashow model and there is no string or zero line of the Higgs field connecting the monopole and adjacent antimonopole.   

The magnetic dipole moments $\mu_m$ for all the MAC configurations are expected to be zero and this is confirmed by the numerical results, that is $\mu_m=0$ for both the SU(2) and U(1) magnetic fields.
The U(1) magnetic field and the SU(2) 't Hooft magnetic field \cite{kn:3} are given by,
\begin{eqnarray}
g^\prime B_i^{U(1)} &=& -\frac{g^\prime}{2}\epsilon^{ijk}G_{jk} = -\epsilon^{ijk}\partial_j \{B_1\sin\theta\}\partial_k\phi \nonumber\\
gB^{SU(2)}_i &=& -\frac{g}{2}\epsilon^{ijk}\hat{F}_{jk} = -\epsilon^{ijk}\partial_j \{gA^{\prime 3}_k\}= \epsilon^{ijk}\partial_j\{g{\cal A}^H_k\} \nonumber\\
&=& -\epsilon^{ijk}\partial_j \{A_1 \sin\theta\}\partial_k\phi
\label{eq.39}
\end{eqnarray}
respectively can be shown by plotting for the respective magnetic field lines. The U(1) magnetic field lines can be shown by drawing the lines of constant of $\{B_1\sin\theta\}$ and the SU(2) t' Hooft magnetic field lines can be shown by drawing the lines of constant of $\{A_1\sin\theta\}$.

\begin{figure}[tbh]
\centering
\hskip-0.1in
\includegraphics[width=5.5in,height=6.0in]{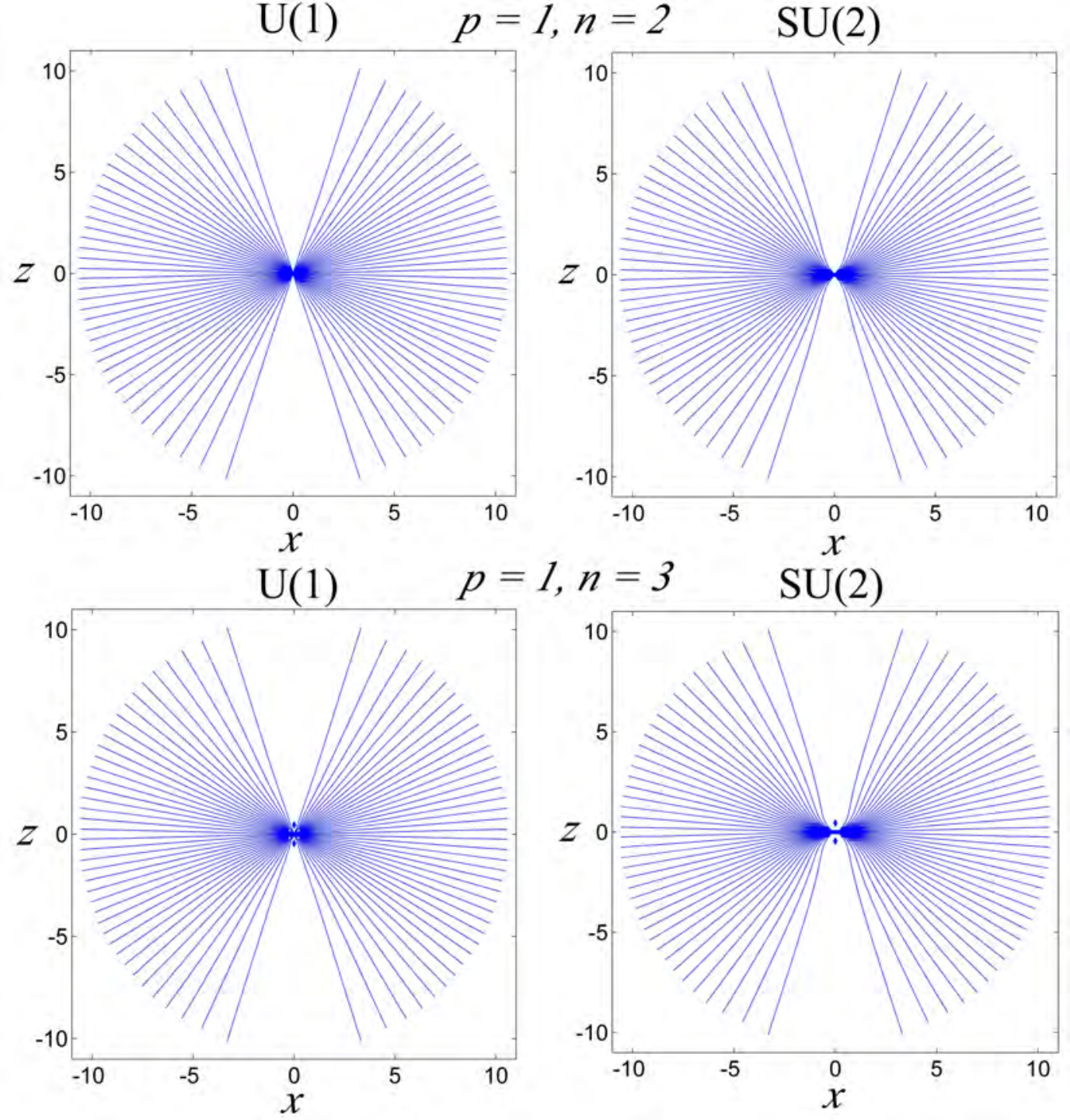} 
\caption{The contour line plot of the U(1) and SU(2) magnetic field lines for the one monopole, $n$-M configurations along the $x$-$z$ plane when $n=2$ and 3. Here $\lambda=\nu=1$ and $\theta_W=\frac{\pi}{4}$.}
\label{fig.2}
\end{figure}

\begin{figure}[tbh]
\centering
\hskip-0.1in
\includegraphics[width=5.5in,height=8.0in]{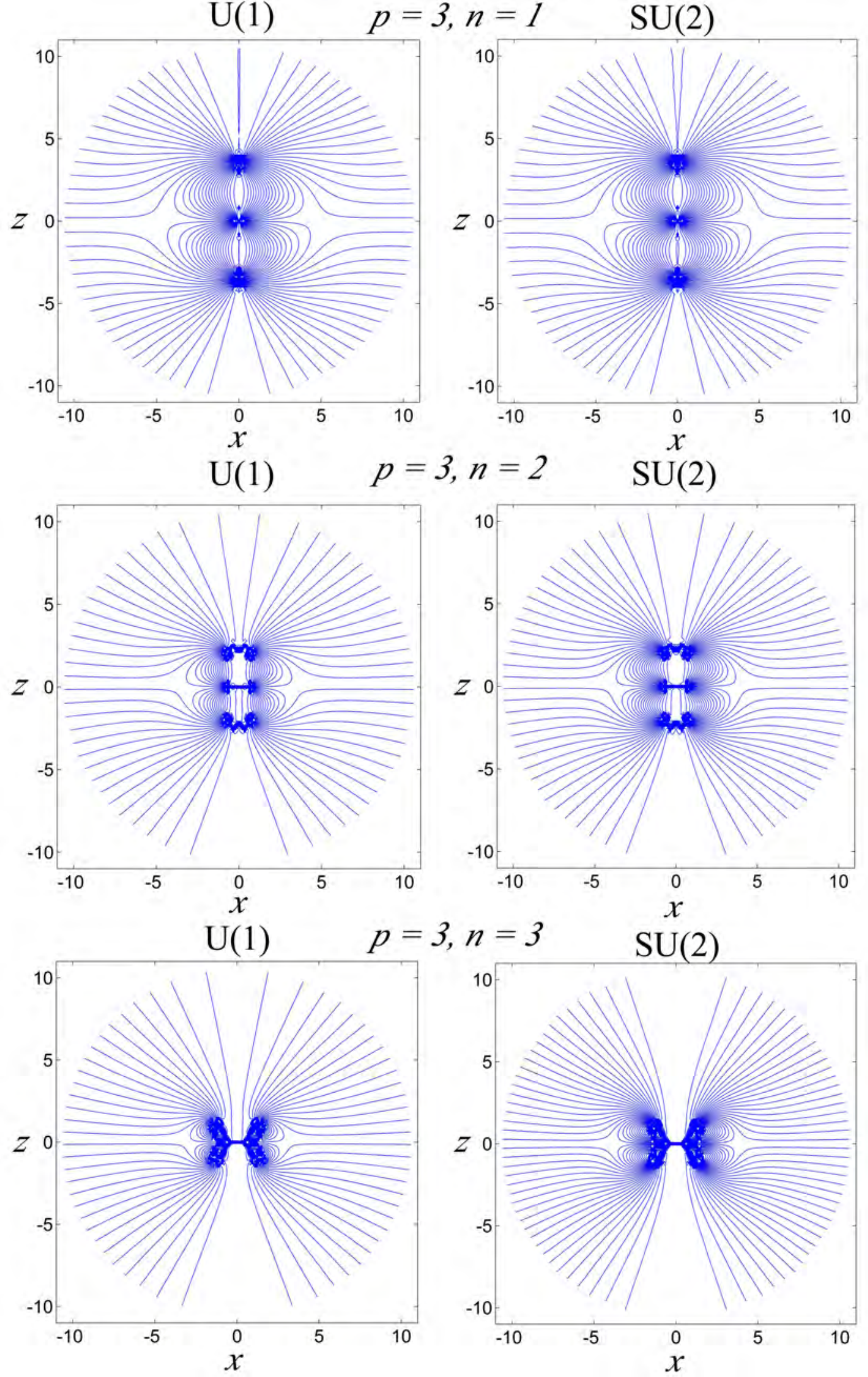} 
\caption{The contour line plot of the U(1) and SU(2) magnetic field lines for the $p=3$, $n$-MAM configurations along the $x$-$z$ plane when $n=1, 2$ and 3. Here $\lambda=\nu=1$ and $\theta_W=\frac{\pi}{4}$.}
\label{fig.3}
\end{figure}

\begin{figure}[tbh]
\centering
\hskip-0.1in
\includegraphics[width=5.5in,height=8.0in]{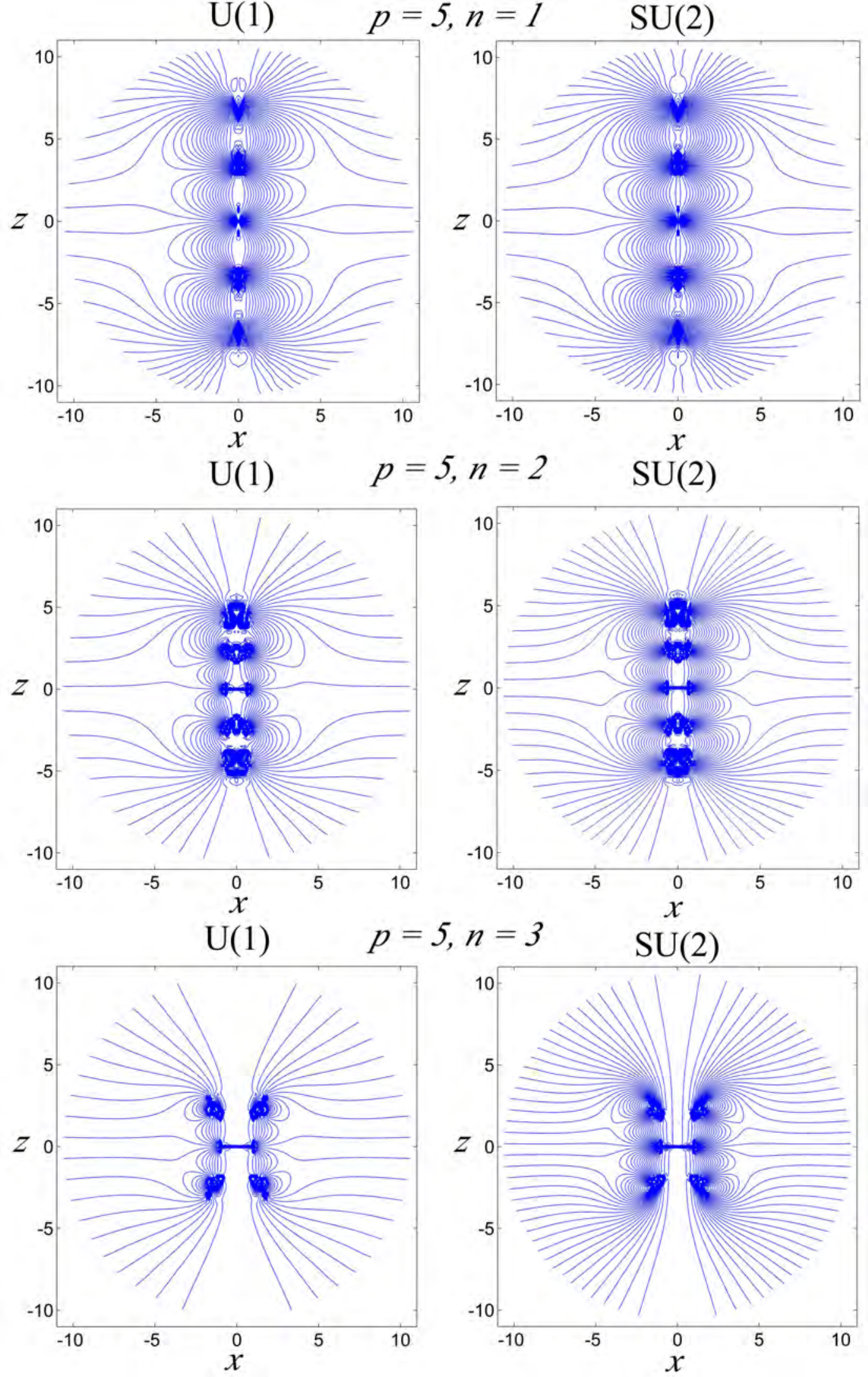} 
\caption{The contour line plot of the U(1) and SU(2) magnetic field lines for the $p=5$, $n$-MAMAM configurations along the $x$-$z$ plane when $n=1, 2$ and 3. Here $\lambda=\nu=1$ and $\theta_W=\frac{\pi}{4}$.}
\label{fig.4}
\end{figure}

The U(1) and SU(2) magnetic field lines for the one monopole configurations are shown in Figure \ref{fig.2} for $\lambda=\nu=1$, $\theta_W=\frac{\pi}{4}$ and when $n=2$ and 3 and they look almost similar. There is a one monopole in both the U(1) field and SU(2) field and hence in the electromagnetic field but the monopole is absent in the neutral ${\cal Z}^0$ field. Figure \ref{fig.3} and \ref{fig.4} show the U(1) and SU(2) magnetic field lines of the three poles $n$-MAM  and five poles $n$-MAMAM configurations respectively for $\lambda=\nu=1$, $\theta_W=\frac{\pi}{4}$ and when $n=1$, 2, and 3. Similar to the one monopole configurations of Figure \ref{fig.2}, the monopoles and vortex-rings found in the SU(2) field are also found in the U(1) field and hence in the electromagnetic field. There are no monopoles or vortex-rings in the neutral ${\cal Z}^0$ field. The monopoles/antimonopoles in the electromagnetic field are Cho-Maison monopoles/antimonopoles with magnetic charges $\pm\frac{4\pi n}{e}$. The vortex-rings however possess zero net magnetic charge.

Similar to the $n$-MAM three poles and $n$-MAMAM five poles configurations of the SU(2) Georgi-Glashow model when the $\phi$-winding number $n=3$, the configurations are transformed into configurations with a $3$-M single pole located at $r=0$ and two vortex-rings of geometric properties $d_{(\rho,z)} = (\rho_i, \pm z_i)$ \cite{kn:15}. In the MAC configurations, the monopoles and vortex-rings are found only in the electromagnetic gauge field, there is no monopole or vortex-ring present in the neutral ${\cal Z}^0$ gauge field.

\begin{figure}[tbh]
\centering
\hskip-0.1in
\includegraphics[width=5.5in,height=8.0in]{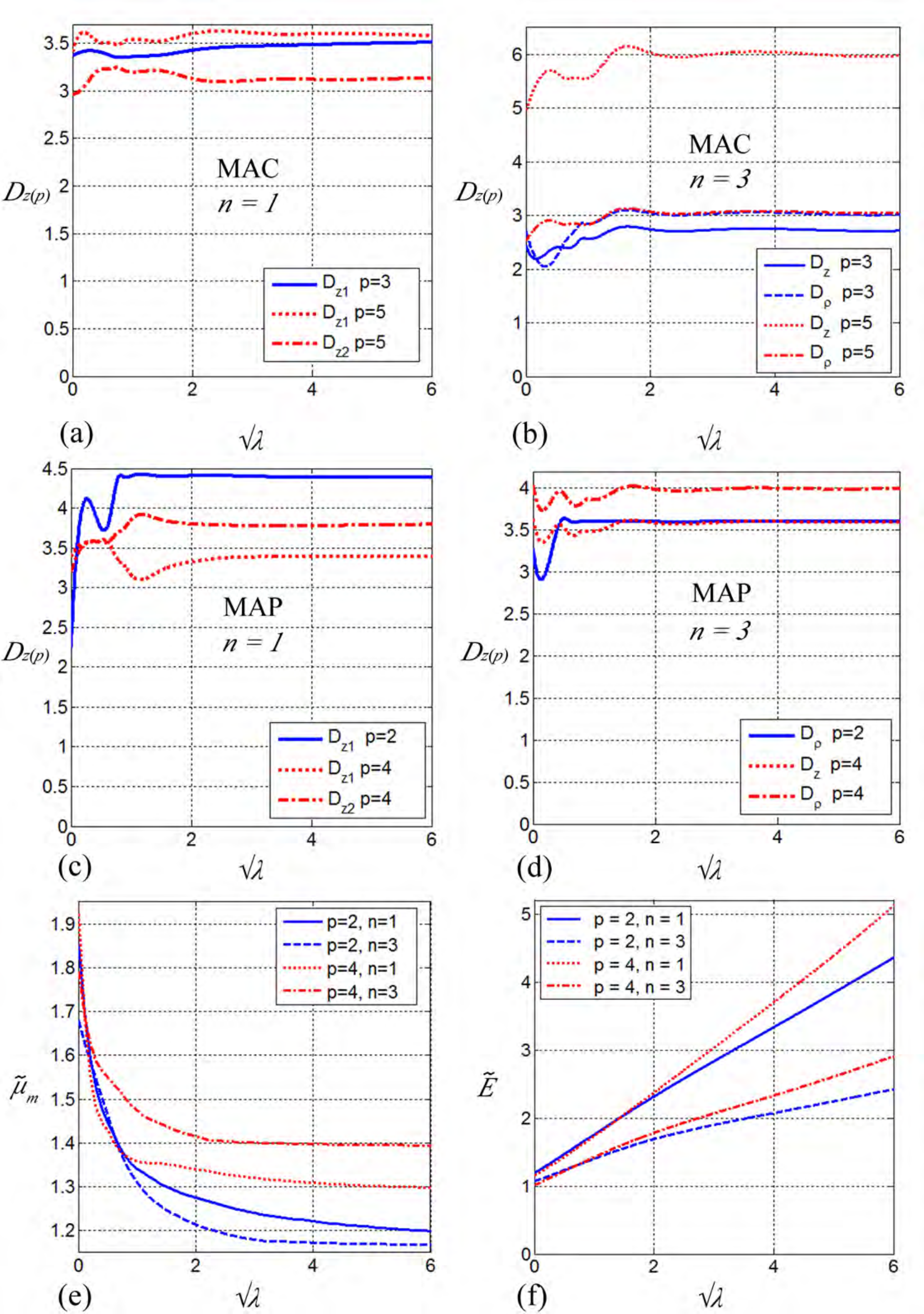} 
\caption{The graphs of $D_{z(\rho)}$  versus $\sqrt{\lambda}$ when (a) $n=1$ and (b) $n=3$, for the $p=3$, 5, MAC configurations and when (c) $n=1$ and (d) $n=3$, for the $p=2$, 4, MAP configurations. (e) The graphs of magnetic dipole moment $\tilde{\mu}_{m}$ versus $\sqrt{\lambda}$ when $n=1$, 3 for the MAP configurations. (f) The graphs of energy $\tilde{E}$ versus $\sqrt{\lambda}$ when $n=1$, 3 for the MAP configurations. Here $\nu=1$ and $\theta_W=\frac{\pi}{4}$.}
\label{fig.5}
\end{figure}

The separation distance $D_{z1}=|\pm z_1|$ of the monopole/antimonopole located at $r=0$ with the adjacent antimonopole/monopole, the separation distance $D_{z2}=|\pm z_2|$ of the chain's outer monopole/antimonopole with the adjacent antimonopole/monopole, the separation distance $D_{z}=2z$ of the two vortex-rings along the $z$-axis, and the diameter $D_\rho=2\rho$ of the vortex-ring, are also plotted versus $\sqrt{\lambda}$ for $0\leq \lambda\leq 40$ when $\nu=1$ and $\theta_W=\frac{\pi}{4}$. The graphs are shown in Figure \ref{fig.5} (a) and (b) when $n=1$ and 3 respectively for the $n$-MAM and $n$-MAMAM configurations. The graphs show that $D_{z1}$, $D_{z2}$, $D_{z}$ and $D_\rho$ become constant when $\sqrt{\lambda}>2$. Hence the separations of monopoles, antimonopoles, and vortex-rings do not change much at large Higgs self-coupling constant or Higgs boson mass.

The total energy of the MAC configurations is infinite due to the presence of point magnetic charges in the U(1) field. The energy density ${\cal E}_0$ blows up at the locations of the point magnetic charge in the U(1) field. Hence the mass of these monopoles can only be estimated to be about 4 to 7 TeV as was done for the one monopole solution when $n=1$ in Ref. \cite{kn:13}.

\begin{table}[tbh]
\begin{center}
\begin{tabular}{|c|c|c|c|}
\hline
$d_{(\rho,z)}$\cellcolor{yellow} & $n=1$\cellcolor{yellow} & $n=2$\cellcolor{yellow} & $n=3$\cellcolor{yellow}  \\ \hline
$p=3$ & $(0,0)$, $(0,\pm 3.375)$ & $(0,0)$, $(0,\pm 2.236)$ & $(0,0)$, $(1.424,\pm 1.283)$   \\ 
      &                          &                           &                               \\ \hline

$p=5$ & $(0,0)$, $(0,\pm 3.559)$, & $(0,0)$, $(0,\pm 2.561)$, & $(0,0)$,  \\ 
      &  $(0,\pm 6.737)$        &  $(0,\pm 4.914)$        & $(1.690,\pm 2.928)$ \\ \hline
\hline
$d_{(\rho,z)}$\cellcolor{yellow} & $n=1$\cellcolor{yellow} & $n=2$\cellcolor{yellow} & $n=3$\cellcolor{yellow}  \\ \hline
$p=2$ & $(0,\pm 2.202)$ & $(0,\pm 1.063)$ & $(1.692,0)$  \\ 
      &                          &                           &                               \\ \hline
$p=4$ & $(0,\pm 1.581)$, & $(0,\pm 1.251)$,  & $(1.606,\pm 1.693)$  \\ 
      &  $(0,\pm 5.435)$ & $(0,\pm 3.166)$   &                      \\ \hline
\end{tabular}
\end{center}
\caption{Table of monopole's and vortex-ring's positions, $d_{(\rho,z)} = (\rho_i, \pm z_i)$, when $\lambda=\nu=1$ and $\theta_W=\frac{\pi}{4}$.}
\label{table.1}
\end{table}

\subsection{MAP/Vortex-ring Configurations}
\label{sect.4.3}

The MAP solutions obtained are the $n$-MA ($p=2$ MAP or 1-MAP) and the $n$-MAMA ($p=4$ MAP or 2-MAP) configurations when the $\phi$-winding number $n=1$ and 2. When $n = 3$, the two poles configuration is transformed into the one vortex-ring configuration and the four poles configuration is transformed into the two vortex-rings configuration just as in the SU(2) Georgi-Glashow model \cite{kn:15}. As in Section \ref{sect.4.2}, the profiles functions, $\psi_1$, $\psi_2$, $R_1$, $R_2$, $\Phi_1$, $\Phi_2$, and $B_1$ for the MAP solutions are obtained numerically and they are all bounded functions of $r$ and $\theta$.

\begin{figure}[tbh]
\centering
\hskip-0.1in
\includegraphics[width=5.5in,height=8.0in]{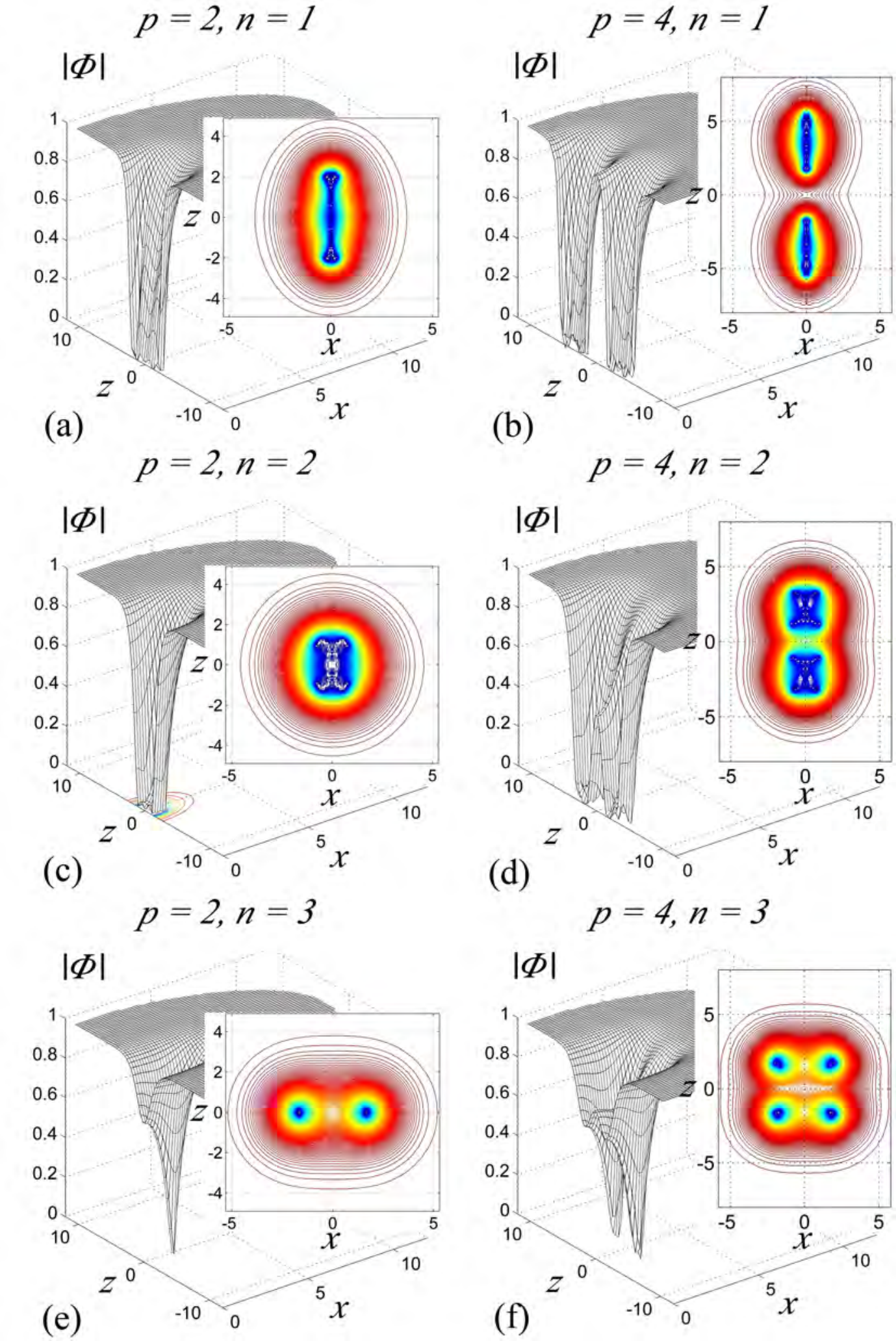} 
\caption{The 3D and contour line plots of the Higgs field modulus $|\Phi|$ along the $x$-$z$ plane for the MAP configurations when (a) $p=2$, $n=1$, (b) $p=4$, $n=1$, (c) $p=2$, $n=2$, (d) $p=4$, $n=2$,  (e) $p=2$, $n=3$, and (f) $p=4$, $n=3$. Here $\lambda=\nu=1$ and $\theta_W=\frac{\pi}{4}$.}
\label{fig.6}
\end{figure}

The 3D and contour line plots of the Higgs field modulus $\Phi$ along the $x$-$z$ plane for the $n$-MA and $n$-MAMA configurations are shown in Figure \ref{fig.6} for $n=1$, 2, 3, $\lambda=\nu=1$ and $\theta_W=\frac{\pi}{4}$. The locations $d_{(\rho,z)}=(\rho_i,\pm z_i)$ of the magnetic monopoles and vortex-rings are read from the zeros of the Higgs field modulus and tabulated in Table \ref{table.1} for the $p=2$ and $p=4$ MAP configurations when $n=1$, 2, and 3. There is however a major difference in the Higgs field modulus here compared to the corresponding MAP solutions of the SU(2) Georgi-Glashow model. The Higgs field modulus is not only zero at the locations of the magnetic monopoles but also vanishes along a line of finite thickness connecting the monopole to the antimonopole of the pair. These observations are not surprising as they are in line with the theoretical predictions of Nambu \cite{kn:5} and the work of others \cite{kn:6} - \cite{kn:9}. 

In the 1-MAP configurations, the separation of the two poles is $D_{z}^{\mbox{\tiny{1-MAP}}}=|2z|$. In the 2-MAP configurations, the separation of the two poles is $D_{z}^{\mbox{\tiny{2-MAP}}}=|z_2-z_1|$ and the separation of the two MAPs is $D_{z}^{\mbox{\tiny{Sp}}}=|2z_1|$. The separation of the vortex-rings is $D_z=|2z|$ along the $z$-axis and the diameter of the vortex-ring is $D_\rho=2\rho$. These values are noted and plotted versus $\sqrt{\lambda}$ for 1-MAP and 2-MAP configurations for $\nu=1$, $\theta_W=\frac{\pi}{4}$ and when $n=1$ and 3 in Figure \ref{fig.5} (c) and (d) respectively. Similar to the MAC configurations in Section \ref{sect.4.2}, $D_{z}^{\mbox{\tiny{1-MAP}}}$, $D_{z}^{\mbox{\tiny{2-MAP}}}$ $D_{z}^{\mbox{\tiny{Sp}}}$, $D_z$ and $D_\rho$ become constant when $\sqrt{\lambda}>2$.

\begin{figure}[tbh]
\centering
\hskip-0.1in
\includegraphics[width=5.5in,height=6.0in]{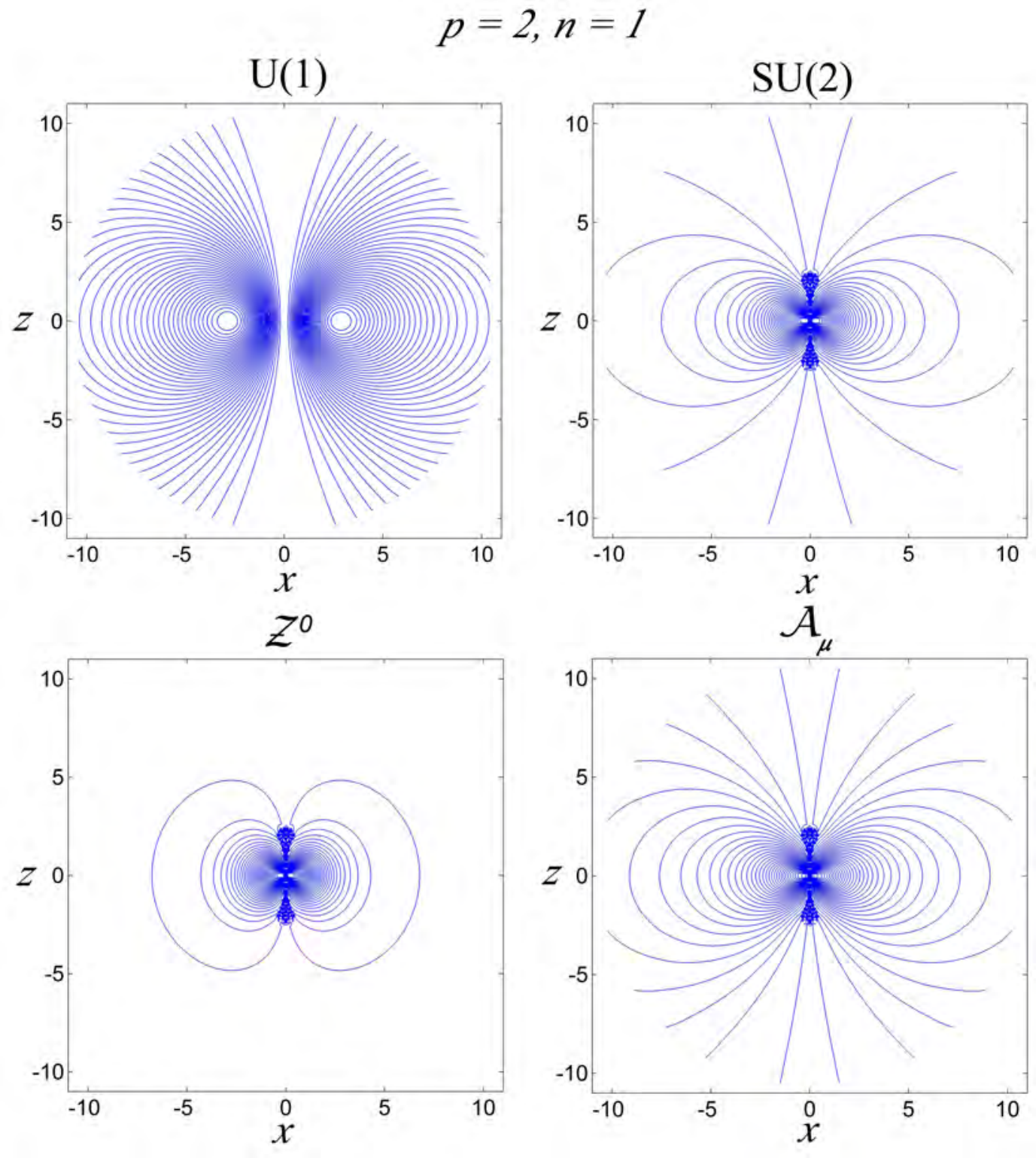} 
\caption{The contour line plot of the U(1), SU(2), neutral ${\cal Z}^0$, and electromagnetic field lines for the $n$-MA configuration along the $x$-$z$ plane when $p=2$, $n=1$. Here $\lambda=\nu=1$ and $\theta_W=\frac{\pi}{4}$.}
\label{fig.7}
\end{figure}

\begin{figure}[tbh]
\centering
\hskip-0.1in
\includegraphics[width=5.5in,height=6.0in]{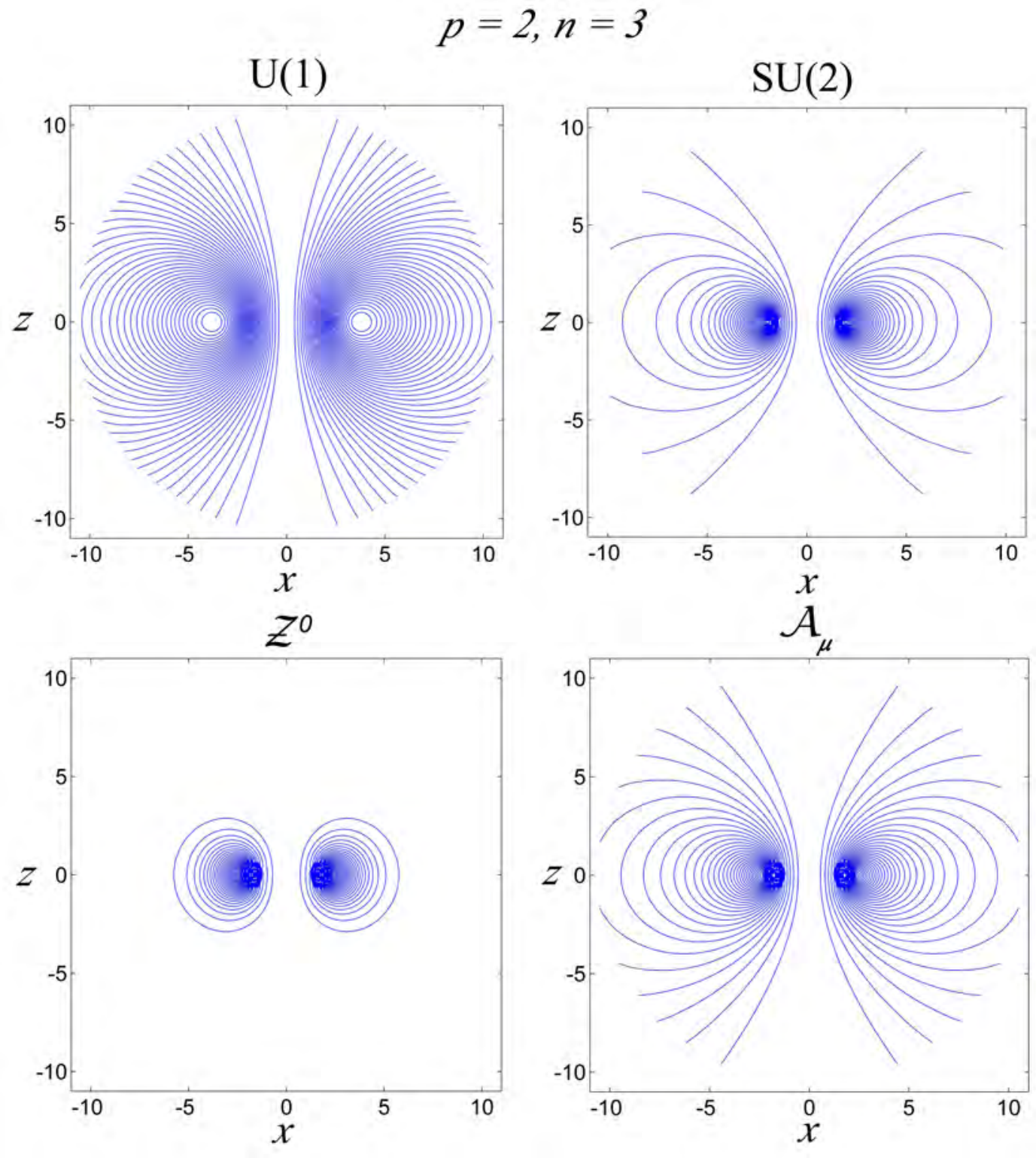} 
\caption{The contour line plot of the U(1), SU(2), neutral ${\cal Z}^0$, and electromagnetic field lines for the $n$-MA configuration along the $x$-$z$ plane when $p=2$, $n=3$. Here $\lambda=\nu=1$ and $\theta_W=\frac{\pi}{4}$.}
\label{fig.8}
\end{figure}

\begin{figure}[tbh]
\centering
\hskip-0.1in
\includegraphics[width=5.5in,height=6.0in]{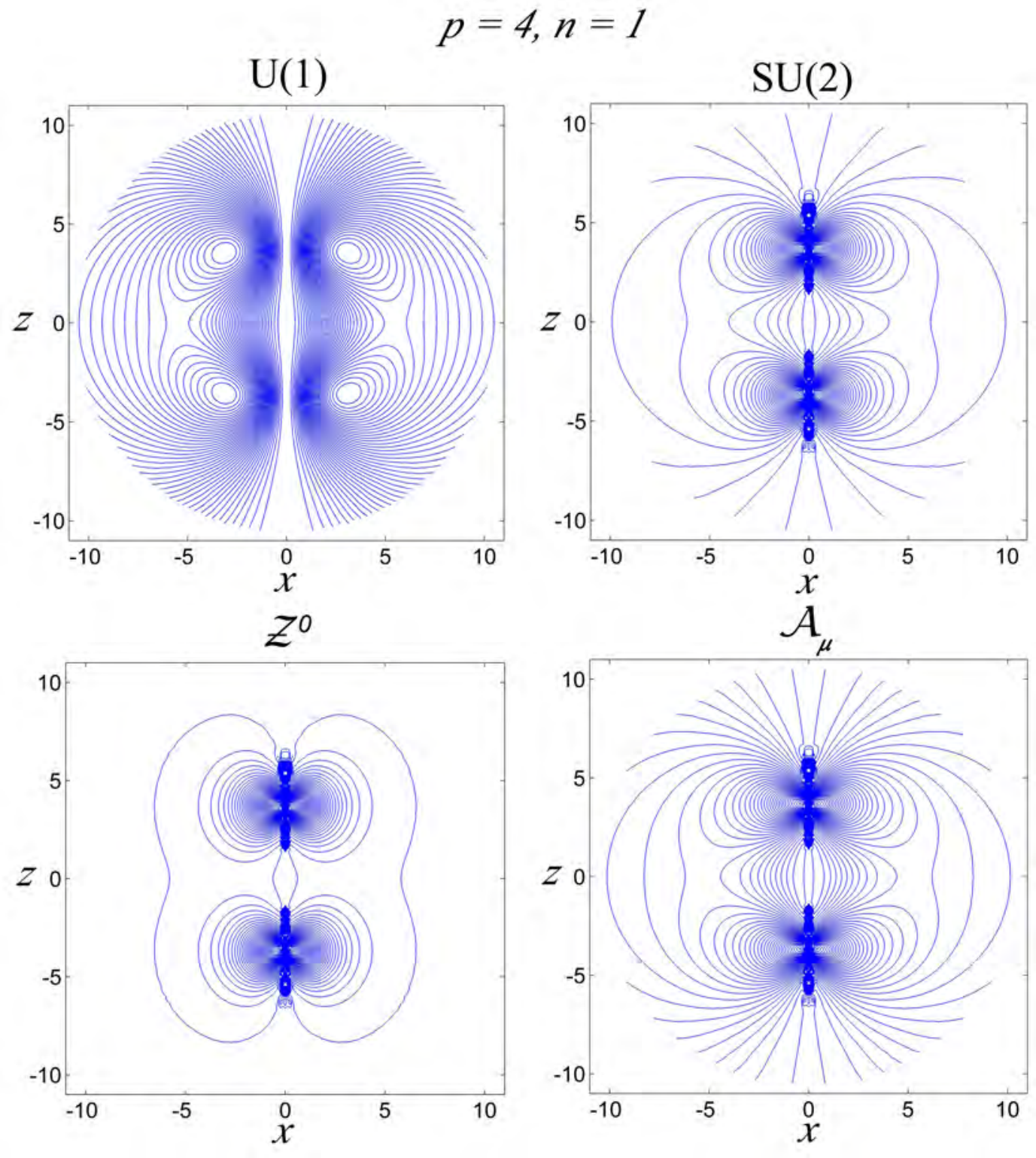} 
\caption{The contour line plot of the U(1), SU(2), neutral ${\cal Z}^0$, and electromagnetic field lines for the $n$-MAMA configuration along the $x$-$z$ plane when $p=4$, $n=1$. Here $\lambda=\nu=1$ and $\theta_W=\frac{\pi}{4}$.}
\label{fig.9}
\end{figure}

\begin{figure}[tbh]
\centering
\hskip-0.1in
\includegraphics[width=5.5in,height=6.0in]{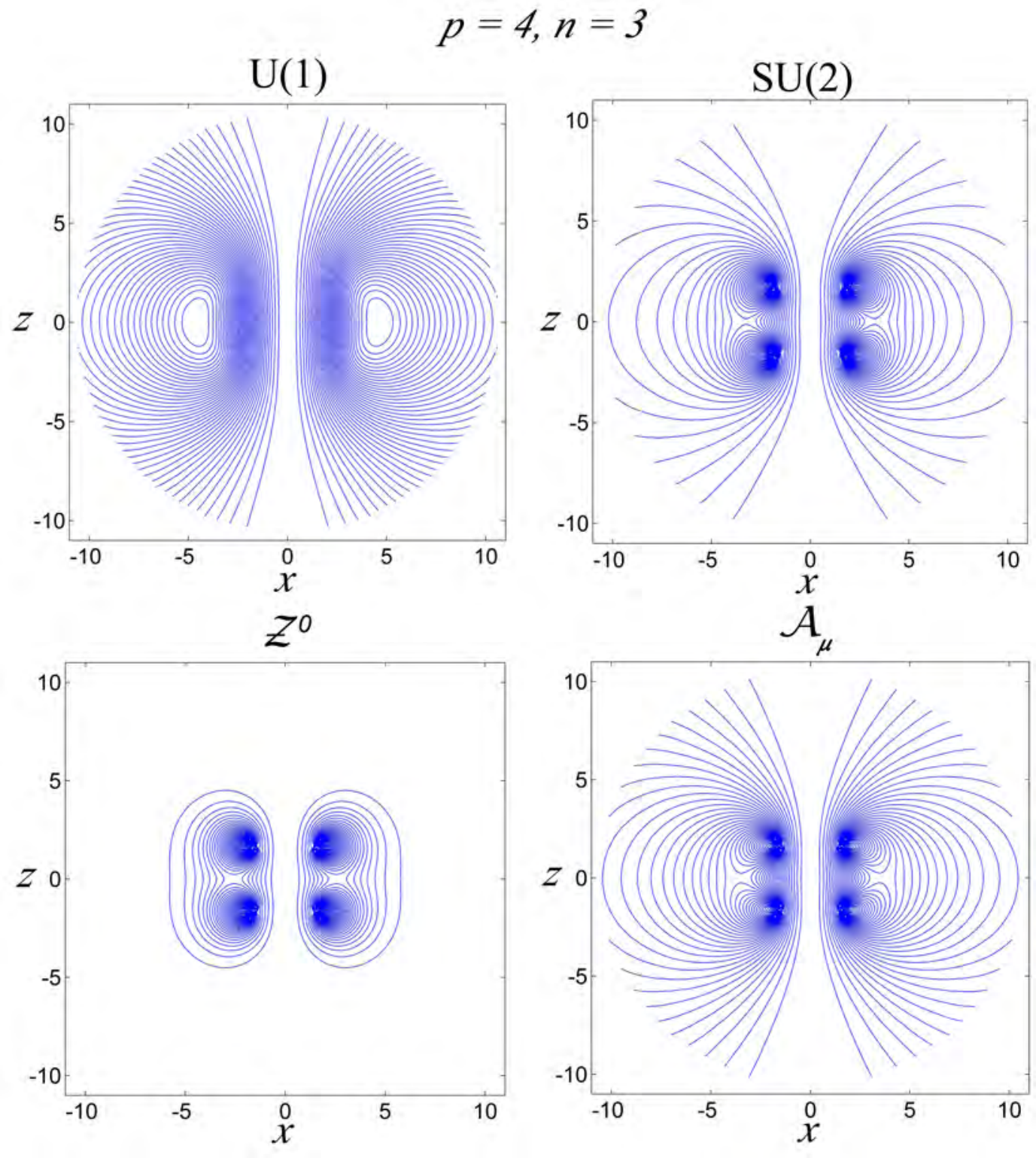} 
\caption{The contour line plot of the U(1), SU(2) Hooft, neutral ${\cal Z}^0$, and electromagnetic field lines for the $n$-MAMA configuration along the $x$-$z$ plane when $p=4$, $n=3$. Here $\lambda=\nu=1$ and $\theta_W=\frac{\pi}{4}$.}
\label{fig.10}
\end{figure}

The magnetic field lines of the Abelian U(1) field, the non-Abelian SU(2) field, the neutral ${\cal Z}^0$ field, and the electromagnetic field when $\lambda=\nu=1$ and the Weinberg angle $\theta_W=\frac{\pi}{4}$ are shown in Figure \ref{fig.7} and \ref{fig.8} for the two poles $n$-MA configurations when $n=1$ and 3 respectively and in Figure \ref{fig.9} and \ref{fig.10} for the four poles $n$-MAMA configurations when $n=1$ and 3 respectively. Unlike the odd poles configurations, there is no monopole or vortex-ring found in the Abelian U(1) field. The monopole and antimonopole pairs and vortex-rings are found only in the SU(2) gauge field. From Eq. (\ref{eq.33}) and (\ref{eq.34}), the magnetic charge of the monopole/antimonopole can be calculated and found to possess fractional magnetic charge of $\pm\frac{4\pi}{e}\sin^2\theta_W$ in the electromagnetic gauge field and fractional magnetic charge of $\pm\frac{4\pi}{e}\sin\theta_W\cos\theta_W$ in the neutral ${\cal Z}^0$ field. When $\theta_W=\frac{\pi}{4}$, $\cos\theta_W=\sin\theta_W=\frac{1}{\sqrt{2}}$ and magnetic charge of the the monopole/antimonopole then possess magnetic charge $\pm\frac{2\pi}{e}$ which is half the magnetic charge of a Cho-Maison monopole both in neutral ${\cal Z}^0$ field and the electromagnetic field.

The SU(2) magnetic field lines of the $n$-MA configuration resembles the magnetic field lines of a bar magnet when $n=1$ and those of the $n$-MAMA configuration resembles the magnetic field lines of a two bar magnets when $n=1$. In the U(1) gauge field, there is totally no magnetic monopole or vortex-ring but only the magnetic flux lines of electric current loops. However, the U(1) magnetic field possesses similar magnetic dipole moment $\mu_m$ as the SU(2) magnetic field as $g^\prime B_i \rightarrow g A^{\prime 3}_i$ at large $r$. In the SU(2) magnetic field, the magnetic dipole moment is due to the monopole-antimonopole pair or vortex-rings but there is no monopole-antimonopole pair or vortex-rings in the U(1) magnetic field, hence in the U(1) field, the magnetic dipole moment must have come from the electric current source loops $\oint{k_i}d^3x$ of Eq. (\ref{eq.23}). There is at least one such current loop in the $n$-MA configurations when $n=1$ and 3 (Figure \ref{fig.7} and \ref{fig.8} respectively) and the $n$-MAMA configuration when $n=3$ (Figure \ref{fig.10}), and two such current loops in the $n$-MAMA configuration when $n=1$ (Figure \ref{fig.9}). 

The magnetic dipole moments $\mu_m$ in unit of $\frac{1}{e}$ for all the even poles configurations are calculated numerically \cite{kn:16}, \cite{kn:20} and the results $\tilde{\mu}_{m}$ are tabulated in Table \ref{table.2} for the $n$-MA ($n=1, 3$) and $n$-MAMA ($n=1, 3$) configurations for $0\leq \lambda\leq 40$. The magnetic dipole moment $\tilde{\mu}_{m}$ which is the magnetic dipole moment $\mu_m$ of the configuration per $n$ and per MAP is plotted versus $\sqrt{\lambda}$ for the $n$-MA and $n$-MAMA configurations when $\nu=1$, $\theta_W=\frac{\pi}{4}$ and $n=1$ and 3 in Figure \ref{fig.5} (e). The graphs of $\tilde{\mu}_{m}$ decrease exponentially fast with increasing $\sqrt{\lambda}$ and the magnetic dipole moments are only additive for a small range of $0.1<\sqrt{\lambda}<0.8$ when the $\tilde{\mu}_{m}$ of the 1-MAP ($n=1, 3$) and 2-MAP ($n=1$) are almost the same. 

The total energy of the MAP configurations is finite due to the fact that there is no point magnetic charge presence in the U(1) field.  The total energy $\tilde{E}$ in unit of $4\pi e$ which is the energy per $n$ per MAP of the configurations is tabulated in Table \ref{table.2} for the 1-MAP ($n=1, 3$) and 2-MAP ($n=1, 3$) configurations for $0\leq \lambda\leq 40$ and also plotted versus $\sqrt{\lambda}$ as shown in Figure \ref{fig.5} (f) when $\nu=1$ and $\theta_W=\frac{\pi}{4}$. We find that the energy $\tilde{E}(\frac{p}{2},n)$ increases approximately linearly with $\sqrt{\lambda}$ when $\sqrt{\lambda}>2$. We also notice that the ratio of the energy is such that $\frac{\tilde{E}(\frac{p}{2},3)}{\tilde{E}(\frac{p}{2},1)}<1$ when $0\leq \sqrt{\lambda}\leq6$ which is consistent with theory.


\begin{table}[tbh]
\begin{center}
\begin{tabular}{cccccccccccc}
\hline 
  \multicolumn{12}{c} {  1-MAP ($n=1$) \cellcolor{yellow} }  \\ 
\hline
$\lambda$ &	0	& 0.05 &	0.1 &	0.5 &	1 & 4 &	8 &	10 &	20 &	30 &	40 \\\hline
$\tilde{E}$ &	1.186 & 1.306 & 1.3527 & 1.581 & 1.751 & 2.317 & 2.734 & 2.905 & 3.572 & 4.088 & 4.520  \\ \hline
$\tilde{\mu}_{m}$	& 1.844 & 1.572 & 1.514 & 1.389 & 1.339 & 1.274 & 1.244 & 1.236 & 1.214 & 1.203 & 1.196 \\ \hline
\hline 
  \multicolumn{12}{c} {  1-MAP ($n=3$) \cellcolor{yellow} }  \\ 
\hline
$\lambda$ &	0	& 0.05 &	0.1 &	0.5 &	1 & 4 &	8 &	10 &	20 &	30 &	40 \\ \hline
$\tilde{E}$ &	1.067 & 1.118 & 1.137 & 1.318 & 1.418 & 1.685 & 1.856 & 1.920 & 2.155 & 2.329 & 2.476 \\ \hline
$\tilde{\mu}_{m}$	& 1.679 & 1.574 & 1.547 & 1.380 & 1.309 & 1.214 & 1.183 & 1.176 & 1.171 & 1.168 & 1.167  \\ \hline
\hline 
  \multicolumn{12}{c} {  2-MAP ($n=1$) \cellcolor{yellow} }  \\ 
\hline
$\lambda$ &	0	& 0.05 &	0.1 &	0.5 &	1 & 4 &	8 &	10 &	20 &	30 &	40 \\ \hline
$\tilde{E}$ &	1.145 & 1.271 & 1.310 &  1.480 & 1.650 & 2.444 & 2.932 & 3.155 & 4.000 & 4.702 & 5.386  \\ \hline
$\tilde{\mu}_{m}$	& 1.801 & 1.525 & 1.483 & 1.384 & 1.358 & 1.340 & 1.322 & 1.318 & 1.305 & 1.300 & 1.296 \\ \hline
\hline 
  \multicolumn{12}{c} {  2-MAP ($n=3$) \cellcolor{yellow} }  \\ 
\hline
$\lambda$ &	0	& 0.05 &	0.1 &	0.5 &	1 & 4 &	8 &	10 &	20 &	30 &	40 \\ \hline
$\tilde{E}$ &	1.009 & 1.086 & 1.102 & 1.310 & 1.453 & 1.772 & 2.012 & 2.102 & 2.459 & 2.746 & 3.000  \\ \hline
$\tilde{\mu}_{m}$	& 1.791 & 1.615 & 1.586 & 1.517 & 1.473 &  1.414 & 1.401 & 1.400 & 1.396 & 1.395 & 1.394 \\ \hline
\end{tabular}
\end{center}
\caption{Table of values of $\tilde{E}$ (total energy per $n$ per MAP) and $\tilde{\mu}_{m}$ (magnetic dipole moment per $n$ per MAP) for the 1-MAP ($n=1$, $n=3$) and 2-MAP ($n=1$, $n=3$) configurations at various values of $0 \leq \lambda \leq 40$. Here $\theta_W = \frac{\pi}{4}$.}
\label{table.2}
\end{table}

\subsection{The Baryon Number}
Since the MAP and vortex-rings solutions of Section \ref{sect.4.3} possess finite energy and zero net topological magnetic charge, the baryon number of these configurations can be calculated using the definition of Ref. \cite{kn:7},
\begin{eqnarray}
Q_B = \frac{g^2}{32\pi^2}\int_{t=t_0}d^3 K^0, ~~
K^0 = \epsilon^{ijk}\left(F^a_{ij}A^a_k - \frac{1}{3}g\epsilon_{abc}A^a_i A^b_j A^c_k\right). 
\label{eq.40}
\end{eqnarray}
The SU(2) gauge potential (\ref{eq.8}) is then gauge transformed into the correct gauge by using the gauge transformation,
\begin{eqnarray}
U_2 &=& \cos\left(\frac{\Theta_2(r)}{2}\right) + i \hat{u}^a_r\sigma^a\sin\left(\frac{\Theta_2(r)}{2}\right),\nonumber\\
&=& \mbox{exp}\left\{\frac{1}{2}i\Theta_2(r)\sigma^a\hat{u}^a_r\right\}, ~~\Theta_2(0)=0 ~\mbox{and} ~\Theta_2(r)|_{r\rightarrow\infty} = -\pi,\nonumber\\
\hat{u}^a_r &=& \sin\left(\frac{p\theta}{2}\right)\cos n\phi ~\delta^a_1 + \sin\left(\frac{p\theta}{2}\right)\sin n\phi ~\delta^a_2 + \cos\left(\frac{p\theta}{2}\right) \delta^a_3,
\label{eq.41}
\end{eqnarray}
where the number of poles $p$ is an even number for the MAP configurations. The gauge transformation $U_2$ rotates the Higgs field direction $\hat{\Phi}^a=-\hat{h}^a$ to $-\delta^a_3$ as $r\rightarrow\infty$ but leaves the Higgs direction unrotated at the origin $r=0$.  The baryon number (\ref{eq.40}) calculated from the gauge transformed potential will give
\begin{eqnarray}
Q_B &=& \frac{n}{4}(1-(-1)^{p/2}), ~p=\mbox{even number}.
\label{eq.42}
\end{eqnarray}
Hence the baryon number for the $n$-MA or 1-MAP configurations is $Q_B = \frac{n}{2}$ and the baryon number for the $n$-MAMA or 2-MAP configurations is $Q_B=0$. Therefore the $n$-MA configuration is a sphaleron and the $n$-MA configuration is a sphaleron and an anti-sphaleron. 

\subsection{The Structure of the Sphaleron}
The structure of the sphaleron has been carefully described by Nambu \cite{kn:5} and Hindmarsh and James \cite{kn:8}. Following along their line of interpretation for the $n$-MA configurations when $n=1$, there is a tube of singularity joining the monopole at $+z_i$ and antimonopole at $-z_i$. Let the SU(2) flux outside the tube and inside the tube be 
\begin{eqnarray}
|\Phi^{\mbox{\tiny out}}_{\mbox{\tiny SU(2)}}| = \frac{4\pi}{g}\eta ~~\mbox{and}~~ |\Phi^{\mbox{\tiny in}}_{\mbox{\tiny SU(2)}}| = \frac{4\pi}{g}\chi, 
\label{eq.43}
\end{eqnarray}
respectively, where $\eta + \chi =1$. According Ref. \cite{kn:5} - \cite{kn:8}, and our numerical result in subsection \ref{sect.4.3}, there is a current loop in the U(1) field. Hence assuming that the U(1) flux that threads through the tube is $\Phi^{\mbox{\tiny in}}_{\mbox{\tiny U(1)}} = \frac{4\pi}{g^\prime}\eta$, then the U(1) flux out side the tube is also $\Phi^{\mbox{\tiny out}}_{\mbox{\tiny U(1)}} = \frac{4\pi}{g^\prime}\eta$. Hence using Eq. (\ref{eq.32}), and assuming that the electromagnetic flux inside the tube is zero, then
\begin{eqnarray}
\Phi^{\mbox{\tiny in}}_{\mbox{\tiny EM}} &=& \cos\theta_W \,\Phi^{\mbox{\tiny in}}_{\mbox{\tiny U(1)}} + \sin\theta_W \,\Phi^{\mbox{\tiny in}}_{\mbox{\tiny SU(2)}} 
= \cos\theta_W\left(\frac{4\pi}{g^\prime}\eta\right) + \sin\theta_W\left(-\frac{4\pi}{g}\chi\right)\nonumber\\
&=& \frac{4\pi}{e}(\cos^2\theta_W \eta - \sin^2\theta_W \chi) = 0,
\label{eq.44}
\end{eqnarray}
implies that $\eta=\sin^2\theta_W$ and $\chi=\cos^2\theta_W$. Then the electromagnetic flux outside the tube coming out from the monopole at $+z_i$ is 
\begin{eqnarray}
\Phi^{\mbox{\tiny out}}_{\mbox{\tiny EM}} &=& \cos\theta_W \,\Phi^{\mbox{\tiny out}}_{\mbox{\tiny U(1)}} + \sin\theta_W \,\Phi^{\mbox{\tiny out}}_{\mbox{\tiny SU(2)}} = \cos\theta_W\left(\frac{4\pi}{g^\prime}\eta\right) + \sin\theta_W\left(\frac{4\pi}{g}\eta\right)\nonumber\\
&=& \frac{4\pi}{e}\eta = \frac{4\pi}{e}\sin^2\theta_W.
\label{eq.45}
\end{eqnarray}
Therefore the magnetic charge of the monopole is $\frac{4\pi}{e}\sin^2\theta_W = \frac{2\pi}{e}$ which is half the magnetic charge of a full Cho-Maison monopole, when the Weinberg angle is $\theta_W=\frac{\pi}{4}$. Similarly, the neutral ${\cal Z}^0$ field flux outside the tube can be calculated to be zero,
\begin{eqnarray}
\Phi^{\mbox{\tiny out}}_{\tiny {\cal Z}^0} &=& -\sin\theta_W\,\Phi^{\mbox{\tiny out}}_{\mbox{\tiny U(1)}} + \cos\theta_W \,\Phi^{\mbox{\tiny out}}_{\mbox{\tiny SU(2)}} = -\sin\theta_W\left(\frac{4\pi}{g^\prime}\eta\right) + \cos\theta_W\left(\frac{4\pi}{g}\eta\right)\nonumber\\
&=& \frac{4\pi}{e}\eta(-\sin\theta_W\cos\theta_W + \cos\theta_W\sin\theta_W) = 0,
\label{eq.46}
\end{eqnarray}
and the neutral ${\cal Z}^0$ field flux inside the tube can be calculated to be
\begin{eqnarray}
\Phi^{\mbox{\tiny in}}_{\tiny {\cal Z}^0} &=& -\sin\theta_W\,\Phi^{\mbox{\tiny in}}_{\mbox{\tiny U(1)}} + \cos\theta_W \,\Phi^{\mbox{\tiny in}}_{\mbox{\tiny SU(2)}} = -\sin\theta_W\left(\frac{4\pi}{g^\prime}\eta\right) + \cos\theta_W\left(-\frac{4\pi}{g}\chi\right)\nonumber\\
&=& -\frac{4\pi}{e}\sin\theta_W\cos\theta_W, 
\label{eq.47}
\end{eqnarray}
where the negative sign of the flux means that it is in the negative $z$ direction.

From our numerical results and calculations, we have confirmed that the $n$-MA configurations, where $n=1, 2, 3$ are sphalerons with baryon number $Q_B=\frac{n}{2}$ and the $n$-MAMA configurations, where $n=1, 2, 3$ are sphaleron-antisphaleron pairs with baryon number $Q_B=0$. Hence the structure of a sphaleron can be a monopole-antimonopole pair, when $n=1, 2$, or a vortex-ring, when $n=3$. There is also an electric current loop circulating around the dipole/ vortex-ring in the $x$-$y$ plane. In the case where the Weinberg angle is $\theta_W=\frac{\pi}{4}$, the monopole and antimonopole charges are $\pm\frac{2\pi}{e}$ respectively which is half the magnetic charge of a Cho-Maison monopole. The U(1) field carries an electric current source loop which contributes to the magnetic dipole moment $\mu_m$ of the U(1) field whereas the SU(2) field carries the monopole-antimonopole pair or vortex-ring which contributes the same magnetic dipole moment $\mu_m$ as the U(1) field. The monopole and antimonopole in the MAP configuration is held by the neutral ${\cal Z}^0$ flux string, $\Phi^{\mbox{\tiny in}}_{\tiny {\cal Z}^0}=-\frac{4\pi}{e}\sin\theta_W\cos\theta_W$. 


\section{Comments}
\label{sect.5}

In summary, we have shown that by using the axially symmetric magnetic ansatz (\ref{eq.8}), we are able to solve numerically for the MAP ($n$-MA and $n$-MAMA) configurations when the parameter $p$ is an even number and the MAC ($n$-M, $n$-MAM and $n$-MAMAM) configurations when $p$ is an odd number although the numerical accuracies of the solutions will decrease with increasing value of parameter $p$ and $\phi$-winding number $n$. The even poles configurations are sphalerons ($n$-MA) and sphalerons-antisphalerons pair ($n$-MAMA), whereas the odd poles configurations are Cho-Maison monopoles-antimonopoles chains ($n$-M, $n$-MAM, $n$-MAMAM).

Our results for the MAP or even poles configurations are in line with the results of Nambu \cite{kn:5} and others \cite{kn:6} - \cite{kn:9}. Our $n$-MA configuration is a sphaleron with finite energy and baryon number $Q_B=\frac{n}{2}$ and that within the sphaleron there is a monopole-antimonopole pair/ vortex-ring together with a loop of electric current circulating the dipole/ vortex-ring. The U(1) field carries the electric current loop and the SU(2) field carries the monopole-antimonopole pair/ vortex-ring \cite{kn:6} - \cite{kn:9}. The monopole-antimonopole pair is also bounded by a flux string of the neutral ${\cal Z}^0$ field \cite{kn:5} - \cite{kn:9}. 
When the Weinberg angle takes the value, $\theta_W=\frac{\pi}{4}$, the monopole and antimonopole are half Cho-Maison monopoles with magnetic charges $\pm\frac{2\pi}{e}$ respectively. This also implies that magnetic dipole moment of the configuration, $\mu_m$, is contributed equally by both the monopole-antimonopole pair/ vortex-ring and the electric current loop. We also note that the $n$-MAMA configurations are a sphalerons-antisphalerons pair with baryon number $Q_B=0$.

Both the MAP configurations in the SU(2) Georgi-Glashow theory \cite{kn:15}, \cite{kn:16} and sphalerons in the Weinberg-Salam theory \cite{kn:7}-\cite{kn:9} are saddle point solutions in their respective theories \cite{kn:21} and are therefore unstable. The neutral ${\cal Z}^0$ field string or flux has been proved to be unstable for a range of the Weinberg angle, $0\leq \sin^2\theta_W<0.8$ when the Higgs boson mass is greater than 24 GeV \cite{kn:22}. Therefore the ${\cal Z}^0$ field string holding the two poles of the MAP together in the sphaleron solution is unstable at $\theta_W=\frac{\pi}{4}$ for values of Higgs mass larger than 24 GeV. We then conclude that both the $n$-MA and $n$-MAMA configurations discussed in this paper are unstable static saddle-point solutions of the Weinberg-Salam theory.

In Ref. \cite{kn:10} and \cite{kn:11}, sphalerons, antisphalerons, and vortex-rings configurations are shown to exist in the SU(2)$\times$U(1) Weinberg-Salam theory by using the same magnetic ansatz (\ref{eq.8}) for the SU(2) gauge field but different Higgs field profile functions. However their numerical results do not reveal the inner structure of the sphaleron. The sphaleron and the sphaleron-antisphaleron pair obtained are just a point particle and a two point particles respectively along the $z$-axis when $n=1$ and 2 and they possess magnetic dipole moment. Hence their results failed to explain the origin of the magnetic dipole moment of the sphaleron. When $n=3$, their one multisphaleron does not give a vortex-ring configuration and their sphalerons-antisphalerons pair becomes a vortex-ring. These results for vortex-ring are different from our results as our one $n$-MA solution gives rise to a vortex-ring and our $n$-MAMA solution gives rise to two vortex-rings configurations when $n=3$. Hence our sphalerons ($n$-MA) and sphalerons-antisphalerons pair ($n$-MAMA) solutions are different from the solutions of Ref. \cite{kn:10} and \cite{kn:11}.

From the numerical results of Ref. \cite{kn:10}, it was given that when $n=1$, the ratios of the magnetic dipole moment $\mu_{(m,n)}$ is  ${\frac{\mu_{(2,1)}}{\mu_{(1,1)}}}|_{\lambda=1}\approx2.0$. They also state that $\frac{\mu_{(m,1)}|_{\lambda=1}}{\mu_{(m,1)}|_{\lambda=0}}\approx 0.75$ where $m=1$ and 2. The parameter $m$ is equivalent to our parameter $\frac{p}{2}$ where they indicate the number of sphalerons in the solution. Our numerical results for magnetic dipole moment $\mu_m(\frac{p}{2},n)$ of the $n$-MA and $n$-MAMA configurations when $n=1$ also give almost the same ratios as that of Ref. \cite{kn:10}, where the ratios ~$\frac{\mu_m(2,1)}{\mu_m(1,1)}|_{\lambda=0}=1.953$, ~$\frac{\mu_m(2,1)}{\mu_m(1,1)}|_{\lambda=1}=2.028$, ~$\frac{\mu_m(1,1)|_{\lambda=1}}{\mu_m(1,1)|_{\lambda=0}}=0.726$~ and $\frac{\mu_m(2,1)|_{\lambda=1}}{\mu_m(2,1)|_{\lambda=0}}=0.754$. 

Again by comparing the ratio of the energy of the sphaleron of Ref. \cite{kn:10} at $\lambda=1$, where ${\frac{E_{(2,1)}}{E_{(1,1)}}}|_{\lambda=1}=1.997\pm 0.003$ with the ratio of the energy of our MAP configurations, ${\frac{E{(2,1)}}{E{(1,1)}}}|_{\lambda=1}=1.884$, we found that the results are also almost the same. Hence the sphaleron and sphaleron-antisphaleron pair solutions of Ref. \cite{kn:10} do possess some similar physical properties as our $n$-MA and $n$-MAMA configurations when $n=1$.

The MAC or odd poles configurations are different from the MAP or even poles configurations as the monopoles and antimonopoles in the MAC configurations are all whole Cho-Maison monopoles and antimonopoles with magnetic charges $\pm\frac{4\pi}{e}$ respectively. The magnetic charges of the monopole and antimonopole do not vary with Weinberg angle. There is no flux string joining a monopole and its adjacent antimonopole in the monopole-antimonopole chain. The U(1) field carries the monopoles and antimonopoles when $n$=1, 2, and vortex-rings when $n$=3. Also the U(1) part of the energy is infinite.

Our MAC configuration is a sequence of monopole-antimonopole chain \{$n$-M, $n$-MAM, $n$-MAMAM, ....\} with the one Cho-Maison monopole solution of Ref. \cite{kn:12} as the first member of the sequence when $n=1$. Although the MAC configurations possess infinite energy, the mass of the monopole can be estimated as was done for the one monopole in Ref. \cite{kn:13} to be about 4 to 10 TeV. Hence by using the axially symmetric magnetic ansatz (\ref{eq.8}), we are able to solve numerically for the whole family of MAC configurations although the numerical accuracy of the solutions will decrease with increasing value of odd number of $p$.

Years ago, Coleman \cite{kn:23} had noted that there is no unique way of representing the electromagnetic field in the region outside the Higgs vacuum at finite value of $r$. One proposal was given by
't Hooft as in equation (\ref{eq.31}) and another was given in equation (4) of Ref. \cite{kn:8}. In our work here, we prefer the definition of 't Hooft as the numerical results obtained for the U(1) gauge potential is very close to the 't Hooft gauge potential that is $g^\prime B_i \approx g A^{\prime 3}_i$ not only at $r\rightarrow \infty$ but for all space for the $n=1$ MAC configurations. Hence the neutral ${\cal Z}^0$ field is almost zero and the electromagnetic field is almost independent of the Weinberg angle $0\leq\theta_W\leq\frac{\pi}{2}$.

The MAP, MAC, and vortex-ring configurations studied here are electrically neutral solutions. Hence further works can be done by introducing electric charges into these configurations. By setting the time component of the gauge field functions, $A_0$ and $B_0$, of the magnetic ansatz (\ref{eq.8}) and (\ref{eq.10}) to be nonvanishing, electric charges will be introduced into the MAP, MAC, and vortex-ring configurations of the SU(2)$\times$U(1) Weinberg-Salam model. This work on the MAC, MAP, and vortex-ring dyons will be reported in the near future.

Further investigation of the MAP configurations are also on the way by varying the Weinberg angle $\theta_W$ from zero to $\frac{\pi}{2}$ and the $\phi$-winding number from $n=1$ to 5. Unlike the MAC configurations, where the monopoles are Cho-Maison monopoles, the monopole and antimonopole of the MAP possess fractional magnetic charges $\pm\frac{4\pi}{e}\sin^2\theta_W$ that are held together by the neutral ${\cal Z}^0$ field flux, $|\Phi^{\mbox{\tiny in}}_{\tiny {\cal Z}^0}| = \frac{4\pi}{e}\sin\theta_W\cos\theta_W$. As $n$ increases from one to two the separation $d_{\mbox{\tiny MAP}}$ of the two poles in the MAP decreases as shown in Table \ref{table.1} for $\lambda=\nu=1$, and $\theta_W=\frac{\pi}{4}$ case. However when $n=3$, the two poles come together to form a vortex-ring instead of annihilating each other. This is because unlike electric charges which are non-topological in nature, magnetic charges are topological in nature as magnetic monopole and antimonopole are topological solitions. It was mentioned in Ref. \cite{kn:8}, that in the limit $d_{\mbox{\tiny MAP}}\rightarrow 0$, where the singular line joining the two poles of the MAP reduces to a point, the vacuum configuration will be recovered rather than the sphaleron which has a long-range dipole field. However this description is not accurate as magnetic monopole and antimonopole cannot annihilate each other and we believe that as $d_{\mbox{\tiny MAP}}\rightarrow 0$, the singular line will reduce to a singular point as in the sphaleron solutions of Ref. \cite{kn:10} and \cite{kn:11} or to a singular vortex-ring as in the $n$-MA and $n$-MAMA configurations presented here when $n=3.$

Another direction of investigation into the MAP configurations is to study the configurations for large values of Higgs self-coupling constant $\lambda$ by looking for higher energy branches of the solutions other than the fundamental branch which exist for all values of $\lambda\geq 0$ as was done for the MAP \cite{kn:24}, \cite{kn:24a} and MAC \cite{kn:25}, \cite{kn:25a} solutions of the SU(2) Georgi-Glashow model. Similar bifurcations and transitions of higher energy branches of solutions may also occur in the Weinberg-Salam model. These investigations will be reported in a later work.

We conclude that the MAC and its vortex-rings solutions can exist both in the SU(2) Georgi-Glashow model as well as in the SU(2)$\times$U(1) Weinberg-Salam model. In the Georgi-Glashow model they all possess finite energy and in the Weinberg-Salam model the energy becomes infinite due to point magnetic charges in the U(1) field. Only the MAP and its vortex-rings configurations with zero net magnetic charge possess finite energy in both the Georgi-Glashow model and the Weinberg-Salam model.  

Another monopole configuration that can exist both in the Georgi-Glashow model \cite{kn:20}, \cite{kn:26}, \cite{kn:27} and the Weinberg-Salam model \cite{kn:28} and possesses finite energy is the half-monopole configuration. The half-monopole configuration in the SU(2) Georgi-Glashow model upon symmetry breaking to the U(1) group  possesses finite energy \cite{kn:26}, \cite{kn:27}. We have found that the half-monopole configuration in the SU(2)$\times$U(1) Weinberg-Salam model also possesses finite energy \cite{kn:28}. 

\section*{Acknowledgements}

The authors would like to thank the Ministry of Science, Technology and Innovation for the ScienceFund Grant (account number: 305/PFIZIK/613613) and the Ministry of Higher Education for the Fundamental Research Grant Scheme (203/PFIZIK/6711354). 


\end{document}